\documentclass[%
 reprint,
 amsmath,amssymb,
 aps, 12pt, one column,
superscriptaddress,
noeprint 
]{revtex4-2}

\usepackage{graphicx}
\usepackage{dcolumn}
\usepackage{bm}

\usepackage{xcolor}
\usepackage{gensymb}
\usepackage{color}
\usepackage{tikz}
\usetikzlibrary{shapes}
\usepackage[colorlinks=False]{hyperref} 

\usepackage{graphicx} 
\usepackage{subcaption}
\usepackage{comment}
\begin{document}

\title{Lagrangian stretching reveals stress topology in viscoelastic flows}


\author{Manish Kumar}
\affiliation{Department of Mechanical Engineering, Purdue University, 585 Purdue Mall, West Lafayette, Indiana 47907 USA}
\author{Jeffrey~S. Guasto}
\affiliation{Department of Mechanical Engineering, Tufts University, 200 College Avenue, Medford, Massachusetts 02155, USA}
\author{Arezoo~M. Ardekani}
\affiliation{Department of Mechanical Engineering, Purdue University, 585 Purdue Mall, West Lafayette, Indiana 47907 USA}


\begin{abstract}
Viscoelastic flows are pervasive in a host of natural and industrial processes, where the emergence of nonlinear and time-dependent dynamics regulate flow resistance, energy consumption, and particulate dispersal. Polymeric stress induced by the advection and stretching of suspended polymers feeds back on the underlying fluid flow, which ultimately dictates the dynamics, instability, and transport properties of viscoelastic fluids. However, direct experimental quantification of the stress field is challenging, and a fundamental understanding of how Lagrangian flow structure regulates the distribution of polymeric stress is lacking. In this work, we show that the topology of the polymeric stress field precisely mirrors the Lagrangian stretching field, where the latter depends solely on flow kinematics. We develop a general analytical expression that directly relates the polymeric stress and stretching in weakly viscoelastic fluids for both nonlinear and unsteady flows, which is also extended to special cases characterized by strong kinematics. Furthermore, numerical simulations reveal a clear correlation between the stress and stretching field topologies for unstable viscoelastic flows across a broad range of geometries. Ultimately, our results establish a connection between the Eulerian stress field and the Lagrangian structure of viscoelastic flows. This work provides a simple framework to determine the topology of polymeric stress directly from readily measurable flow field data and lays the foundation for directly linking the polymeric stress to flow transport properties.
\end{abstract}


\maketitle


\section*{Introduction}

The stretching of long-chain polymers in flow imparts viscoelastic properties to fluids, which impact diverse industrial, geophysical, and biological applications \cite{sorbie2013,Tang2013,Thiebaud2014,Kumar2022review}.
Viscoelasticity leads to increased flow resistance in enhanced oil recovery, polymer processing, and microbial mining \cite{Browne2021,Denn2004,Stoodley2005}, and it enhances fluid and particulate transport in targeted drug delivery and reproduction \cite{Walkama2020,Haward2021,Jacob2017}. 
Extensional flow components simultaneously stretch and advect polymeric chains, which creates large and inhomogeneously distributed polymeric stress~\cite{Wagner2016}. 
\textcolor{black}{Viscoelastic instabilities occur~\cite{Larson1992,Pakdel1996}, when elastic stresses dominate viscous stresses, and manifest symmetry breaking \cite{Arratia2006}, time-dependent flow~\cite{Groisman2000}, and enhanced mixing~\cite{Groisman2001}. 
The onset of these phenomena are captured by the Weissenberg
number ({\rm Wi}), representing the ratio of elastic to viscous stress: ${\rm Wi} = \lambda \dot{\gamma}$, where $\lambda$ and $\dot{\gamma}$ are the polymeric relaxation time and deformation rate, respectively.}
Importantly, the topology of the polymeric stress field has been shown to regulate flow structure, whereby streaks of high polymeric stress lead to separation and act as a barrier to flow~\cite{Kumar2021multistability,Kumar2021tristability}. 
\textcolor{black}{Determining the topology of the stress field and its relationship to flow kinematics are fundamental to understanding and predicting dynamic flow patterns and ultimately, material transport in complex flows.}

Direct optical measurements of the stress field and polymer deformation in viscoelastic flows are challenging \cite{Moss2010,Sun2016}. 
Flow-induced birefringence measurements can provide spatially resolved stress fields, but they require highly specialized imaging instruments \cite{fuller1995optical,murphy2002fundamentals}.
Furthermore, large stress-optical coefficients are difficult to achieve for polymeric solutions \cite{Ober2011,Sun2016} and the linear stress-optical rule is not applicable at high stress \cite{Sun2016,Haward2016}. 
Individual polymer stretching measurements \cite{Smith1999,Kawale2017} are possible, but they require single molecule imaging sensitivity, do not provide whole-field information, and are limited to relatively slow flows.
However, recognizing that the polymeric stress distribution is inherently coupled to polymer advection and deformation through flow kinematics suggests that a Lagrangian analysis of viscoelastic flows can provide direct insight into the structure of the polymeric stress field.

The Lagrangian stretching field is a type of Lagrangian coherent structure (LCS) \cite{Haller2001,Voth2002,Haller2015} that has found numerous applications in geophysical flows~\cite{Haller2015}, active and passive particle transport~\cite{Parsa2011,Dehkharghani2019}, and chemical reacting flows, but its use in non-Newtonian flows has been limited~\cite{Arratia2005}.
The stretching field quantifies the relative deformation of fluid elements in flow, but unlike the polymeric stress, it is easily computed from readily measurable velocity fields~\cite{Haller2015}.
Therefore, in this work, we determine the relationship between the polymeric stress and the Lagrangian stretching field for a broad range of viscoelastic fluid flows.
In the limit of small Weissenberg number (${\rm Wi} \ll 1$), theoretical analysis yields a general analytical expression that directly relates the trace of the polymeric stress tensor to the stretching field, \textcolor{black}{which applies even in unsteady and nonlinear flows and is extended to special cases \cite{Noll1962} exhibiting strong kinematics ($\rm Wi \gg 1$). 
Further, numerical simulations at large Weissenberg number (${\rm Wi}\gtrsim 1$) demonstrate the strong correlation between the stretching and stress fields in nontrivial geometries and highly time-dependent, chaotic flows. Taken together, these results fundamentally establish how the Lagrangian flow structure underpins the Eulerian stress distribution in viscoelastic flows.
They also provide a novel framework to determine the polymeric stress field topology for arbitrary flows, 
which will ultimately give new insights into the onset of viscoelastic instabilities and their transport properties.}

\renewcommand{\arraystretch}{1.7}
 \begin{table*}
 \caption{ \label{linear_flows} Analytical polymeric stress and stretching fields for small Weissenberg number (${\rm Wi} \ll 1$) linear flows. Weissenberg numbers ($\rm{Wi}$) for extensional and shear flows are defined as $\rm{Wi}=\dot{\epsilon} \lambda$ and $\rm{Wi}=\dot{\gamma} \lambda$, respectively. Stretching fields ($S$) are determined exactly and also shown in terms of the Taylor expansion up to $O(\rm{Wi}^2)$ with the remaining terms indicated by H.O.T.}
 \begin{center}
\begin{tabular}{|p{2cm}|p{2.8cm}|p{3cm}|p{8cm}|} 
 \hline
 \textbf{Flow type} & \textbf{Velocity field} ($\mathbf{u}$) & \textbf{Stress field} (${\rm tr}(\boldsymbol{\tau_p})$) & \textbf{Stretching field} ($S$)  \\
 \hline
 Extensional flow & $u=\dot{\epsilon} x$,~~~~$v=-\dot{\epsilon} y$ & ${\rm tr}(\boldsymbol{\tau_p})=\frac{8(b_{11}-b_2)}{\lambda^2}\rm{Wi}^2$ & $S^2=e^{2\rm{Wi}}=1+2\rm{Wi}+2\rm{Wi}^2+H.O.T.$  \\
 \hline
 
  Simple shear flow & $u=\dot{\gamma} y$,~~~~$v=0$ & ${\rm tr}(\boldsymbol{\tau_p})=\frac{2(b_{11}-b_2)}{\lambda^2}\rm{Wi}^2$ & $S^2=1+\frac{1}{2}\rm{Wi}^2+\rm{Wi} \left(1+\frac{1}{4}\rm{Wi}^2 \right)^{1/2}=1+\rm{Wi}+\frac{1}{2}\rm{Wi}^2+H.O.T.$  \\
 \hline
 
  Rotational flow & $u=-\Omega y$,~~$v=\Omega x$ & ${\rm tr}(\boldsymbol{\tau_p})=0$ & $S^2=1$  \\
 \hline
 
 \end{tabular}
\end{center}
\end{table*}

\section*{Results and Discussion}

\subsection*{\textcolor{black}{Lagrangian stretching and polymeric stress fields for weakly viscoelastic flows}}
\textcolor{black}{In the limit of small Weissenberg number (${\rm Wi} \ll 1$), we first derive an analytical relationship between the trace of the polymeric stress tensor and the Lagrangian stretching field for simple, linear flows.}
The Lagrangian stretching field quantifies the relative elongation of a fluid element during advection and deformation in flow over a fixed time interval~\cite{Voth2002}. 
To determine the stretching field, material lines in the velocity field, $\mathbf{u}$, are first obtained by integrating $\frac{d\mathbf{x}}{dt} =\mathbf{u}(\mathbf{x}, t)$.
The solution is denoted as the flow map, $\mathbf{\Phi} = \mathbf{x}(t_1, \mathbf{x}_0, t_0)$, which provides a mapping between the initial, $\bold{x_0}$, and final positions of fluid particles due to advection between times $t_0$ and $t_1$.
The Lagrangian history of fluid particle deformation is encoded in the gradients of the flow map and represented by the right Cauchy-Green strain tensor
\begin{equation}\label{c_r}
\bold{C}_R=(\nabla \boldsymbol{\Phi})^\intercal \nabla \boldsymbol{\Phi},
\end{equation}
which is symmetric.
The Lagrangian stretching field, $S(\mathbf{x},t)$, is defined as the square root of the largest eigenvalue of $\bold{C}_R$ \cite{Voth2002}, and the corresponding eigenvector indicates the direction of stretching \cite{Parsa2011}. 
The stretching field is calculated analytically for simple flow fields and numerically for simulated or measured flows, where the time interval, $\Delta t=t_1-t_0$, is chosen based on the natural flow time scale.
For viscoelastic flows, the local polymeric stress at a particular time depends directly on the accrued stretching of the polymeric chains over the course of their relaxation time. 
Thus, to develop a correlation between the polymeric stress and Lagrangian stretching fields, we examine $S$ over the time interval $\Delta t=\lambda$ \textcolor{black}{(unless specified otherwise)}, which represents the relevant time scales for both polymer stretching and relaxation.

\textcolor{black}{In the case of ${\rm Wi} \ll 1$}, various models for viscoelastic fluids converge to the second-order fluid model~\cite{bird1987dynamics_vol1}, and the polymeric stress tensor is given as
\begin{equation}\label{tau_p}
\boldsymbol{\tau_p}=b_1 \boldsymbol{\gamma_{(1)}}+b_2 \boldsymbol{\gamma_{(2)}}+b_{11} \{\boldsymbol{\gamma_{(1)}} \cdot \boldsymbol{\gamma_{(1)}}\},
\end{equation}
where $b_1$ represents the viscosity, and $b_2$ and $b_{11}$ are the first and second normal stress differences, respectively. For the second-order fluid model, the polymeric relaxation time can be given as $\lambda=-b_2/b_1$ (\textit{SI Appendix}, Table~S1) \cite{bird1987dynamics_vol1}.
For weakly viscoelastic fluids, the stress tensor is calculated using the Newtonian velocity field via Giesekus's theorem \cite{bird1987dynamics_vol1}.
$\boldsymbol{\gamma_{(1)}}=\nabla \textbf{u}+(\nabla \textbf{u})^\intercal$ is the strain rate tensor, and its higher-order derivatives are obtained from the following relationship
\begin{equation}\label{gamma}
\boldsymbol{\gamma_{(n+1)}}=\frac{D \boldsymbol{\gamma_n}}{Dt} -\{(\nabla \textbf{u})^\intercal \cdot \boldsymbol{\gamma_n}+ \boldsymbol{\gamma_n} \cdot (\nabla \textbf{u})\},
\end{equation}
where $\frac{D}{Dt}=\frac{\partial}{\partial t}+\bold{u} \cdot \nabla $ is the material derivative.
For linear extensional, shear, and rotational flows, the trace of the stress tensor, ${\rm tr}(\boldsymbol{\tau_p})$, and the stretching field, $S$, were both calculated analytically and are both spatially uniform due to constant $\nabla \mathbf{u}$ and $\nabla \mathbf{\Phi}$, respectively.
The results are summarized in Table~\ref{linear_flows}.
In the special case of uniform (rigid body) motion, for example in rotational flow, the trace of the stress tensor is ${\rm tr}(\boldsymbol{\tau_p})=0$ due to a lack of fluid deformation.
Likewise, $S=1$ as it represents the relative elongation of a fluid element, and the net stretching is $S-1$.
For linear flows (Table \ref{linear_flows}), ${\rm tr}(\boldsymbol{\tau_p})$ and $S$ satisfy the following equation for ${\rm Wi} \ll 1$:
\begin{equation}\label{stretching_stress_relation}
{\rm tr}(\boldsymbol{\tau_p})= \frac{2(b_{11}-b_2)}{\lambda^2}(S^2-1)^2.
\end{equation}
The stretching field has the form $S=1+O(\rm{Wi})$.
Therefore, any power of $S$ can be written as $S^n=1+n O(\rm{Wi})$ to leading order, which gives the general equation:
\begin{equation}\label{stretching_stress_relation_general}
{\rm tr}(\boldsymbol{\tau_p})= \frac{8}{n^2}\frac{(b_{11}-b_2)}{\lambda^2}(S^n-1)^2,
\end{equation}
where $n \neq 0$.
The simplicity of this result suggests that the stretching and stress fields are intrinsically linked.

\subsection*{\textcolor{black}{Extension to non-linear flows with weakly viscoelastic fluids}}
The simple linear flows explored above provide important insights into the relationship between homogeneous polymer stress and stretching fields, but whether such relationships (Eq.~\ref{stretching_stress_relation_general}) hold for topologically complex flows with spatially varying velocity gradients remains to be determined.
Therefore, we derive ${\rm tr}(\boldsymbol{\tau_p})$ and $S$ for a series of spatially nonlinear flows at ${\rm Wi} \ll 1$. 
In a Poiseuille flow through a channel of height $2H$ with center-line flow speed $U_0$ and velocity field components, $u=U_0[1-(y/H)^2]$ and $v=0$, ${\rm tr}(\boldsymbol{\tau_p})$ and $S$ are given by:
\begin{equation}\label{tau_pois}
{\rm tr}(\boldsymbol{\tau_p})=\frac{8(b_{11}-b_2)}{\lambda^2}\Bar{y}^2{\rm Wi}^2, 
\end{equation}
\begin{align}\label{s_pois}
S^2 &=1+  2\Bar{y}^2{\rm Wi}^2+ 2\Bar{y} {\rm Wi} \left(1+\Bar{y}^2{\rm Wi}^2 \right)^{1/2} \\ 
&= 1+2\Bar{y}{\rm Wi}+2\Bar{y}^2{\rm Wi}^2+H.O.T.,
\end{align}
where ${\rm Wi}=U_0\lambda/H$ and $\Bar{y}=\lvert y \rvert/H$. Next, we consider a quadratic extensional flow with $u=\dot{\epsilon_1} xy$ and  $v=-\frac{1}{2}\dot{\epsilon_1} y^2$. The resulting stress and stretching were found to be:
\begin{equation}\label{tau_ext_quad}
{\rm tr}(\boldsymbol{\tau_p})=\frac{2(b_{11}-b_2)}{\lambda^2}(\Bar{x}^2+4\Bar{y}^2){\rm Wi}^2,
\end{equation}
\begin{align}\label{s_ext_quad}
S^2=1&+\left(\Bar{x}^2+4\Bar{y}^2\right)^{1/2}{\rm Wi} \\ &+\frac{1}{2}\left\{\left(\Bar{x}^2+4\Bar{y}^2\right)+\left( \frac{2\Bar{y}^3-\Bar{y}\Bar{x}^2}{\sqrt{\Bar{x}^2+4\Bar{y}^2}}\right)\right\}{\rm Wi}^2+H.O.T.,
\end{align}
where ${\rm Wi}=\dot{\epsilon_1}L_c\lambda$, $\Bar{x}=x/L_c$, $\Bar{y}=y/L_c$, and $L_c$ is the characteristic length scale of flow. 
The two quadratic flow fields examined above exhibit spatially nonuniform stretching and stress fields. 
However, the relationship between ${\rm tr}(\boldsymbol{\tau_p})$ and $S$ derived in Eq.~\ref{stretching_stress_relation_general} for linear flows also holds for nonlinear flows. 
As further validation, we consider the quartic extensional velocity field $u=\frac{1}{2}\dot{\epsilon_2} x^2y^2$ and $v=-\frac{1}{3}\dot{\epsilon_2} xy^3$, and analytically derive ${\rm tr}(\boldsymbol{\tau_p})$ and $S$:
\begin{equation}\label{tau_quart}
{\rm tr}(\boldsymbol{\tau_p})=\frac{2(b_{11}-b_2)}{\lambda^2}\left(\Bar{x}^4\Bar{y}^2+\frac{10}{3}\Bar{x}^2\Bar{y}^4+\frac{1}{9}\Bar{y}^6\right){\rm Wi}^2, 
\end{equation}
\begin{align}\label{s_ext_quart}
S^2=1 &+ \left(\Bar{x}^4\Bar{y}^2+\frac{10}{3}\Bar{x}^2\Bar{y}^4+\frac{1}{9}\Bar{y}^6\right)^{1/2}{\rm Wi} \\ & +\frac{1}{2} \left\lbrace \begin{array}{l}
\left(\Bar{x}^4\Bar{y}^2+\frac{10}{3}\Bar{x}^2\Bar{y}^4+\frac{1}{9}\Bar{y}^6\right)+\\
\frac{1}{3}\left( \frac{-6\Bar{x}^5\Bar{y}^4+5\Bar{x}^3\Bar{y}^6+\Bar{x}\Bar{y}^8}{\sqrt{9\Bar{x}^4\Bar{y}^2+30\Bar{x}^2\Bar{y}^4+\Bar{y}^6}}\right)
\end{array}\right\rbrace
{\rm Wi}^2+H.O.T.,
\end{align}
where ${\rm Wi}=\dot{\epsilon_2}L_c^3\lambda$.
Strikingly, the expression established in Eq.~\ref{stretching_stress_relation_general} also holds for the quartic extensional flow.

\subsection*{\textcolor{black}{Extension to weakly unsteady viscoelastic flows}} 
\textcolor{black}{
To expand the applicability of the relationship between polymeric stress and stretching, we next extend our analysis to time-dependent flows.
Viscoelastic flows are fully characterized not just by $\rm{Wi}$, but also by the dimensionless Deborah number (De$=\lambda/T$). 
The latter is a measure of unsteadiness and corresponds to the ratio of polymeric relaxation time ($\lambda$) to the characteristic time scale of the flow ($T$) \cite{Dealy2010}. 
For weakly viscoelastic ($\rm{Wi \ll 1}$) and weakly unsteady ($\rm{De \ll 1}$) flows, we consider a time-dependent perturbation to a velocity field, $\boldsymbol{u}=\boldsymbol{u_0}[1+ {\rm De} \alpha(t)]$, where $\boldsymbol{u_0}$ is a steady linear or nonlinear flow (e.g. explored in the previous sections), and $\alpha(t)$ is an arbitrary time-dependent function. 
Under these conditions, the ordered fluid model (Eq. \ref{tau_p}) is still applicable \cite{bird1987dynamics_vol1,Ewoldt2017}, and we find that the instantaneous stress and stretching fields (i.e., $t=t_0$) satisfy the following relationship (see \textit{SI Appendix}):
\begin{equation}\label{stretching_stress_relation_unsteady}
{\rm tr}(\boldsymbol{\tau_p})= \frac{8}{n^2}\frac{(b_{11}-b_2)}{\lambda^2}g(t_0)(S^n-1)^2,
\end{equation}
where
\begin{equation}\label{order_unsteady_g}
g(t_0)=\left [\frac{1+{\rm De}\alpha(t_0)}{1+{\rm De}\beta(t_0)} \right ]^2,
\end{equation}
and $\beta(t_0)=-\frac{1}{\lambda}\int_{t_0}^{t_0-\lambda} \alpha (t) dt$.
}

\textcolor{black}{
In the special case of fluid \textit{motions with constant stretch history} (MWCSH)~\cite{Noll1962,Huilgol1969,Huilgol1976,HUILGOL1971} -- for example linear and Poiseuille flows -- the stress tensor can be obtained using the first three kinematic tensors ($\boldsymbol{\gamma_{(1)}}, \boldsymbol{\gamma_{(2)}}$, and $\boldsymbol{\gamma_{(3)}}$) \cite{Noll1958}. 
For $\rm{Wi<1}$, the leading order term (i.e., $O(\rm{Wi})$) of the stretching field still dominates.
Thus, for weakly modulated MWCSH flows ($\rm{ De \ll 1}$), the relationship between the stress and stretching fields at $\rm{Wi < 1}$ is described by Eq.~\ref{stretching_stress_relation_unsteady} (see \textit{SI Appendix}), where specifically
\begin{equation}\label{mwcsh_unsteady_g}
g(t_0)=\frac{\left[\{1+{\rm De}\alpha(t_0)\}^2+\frac{2b_{12}-3b_3}{\lambda(b_{11}-b_2)}\{1+{\rm De}\alpha(t_0)\}{\rm De}\zeta(t_0)\right]}{\left[1+{\rm De}\beta(t_0)\right]^2},
\end{equation}
$\zeta(t_0)=\lambda \alpha^{\prime}(t_0)$, and $b_3$ and $b_{12}$ are the constants associated with the third order kinematic tensors.
We note that in the limit of $\rm{De} \to 0$, $g(t_0) \to 1$ and Eq.~\ref{stretching_stress_relation_unsteady} converges to Eq.~\ref{stretching_stress_relation_general} for steady flows. 
Furthermore, the stress fields obtained for flows undergoing MWCSH are also applicable at $\rm{Wi \gtrsim 1}$. 
For strong kinematics ($\rm{Wi \gg 1}$), the highest order (i.e. $O(\rm{Wi^2})$) term of the stretching field dominates in shear flows. 
Therefore, the relationship between the stress and stretching fields at $\rm{Wi \gg 1}$ for weakly modulated ($\rm{De \ll 1}$) homogeneous (simple shear) and non-homogeneous (Poiseuille) shear flows is: 
\begin{equation}\label{stretching_stress_relation_strong_flow_shear}
{\rm tr}(\boldsymbol{\tau_p})=2\frac{(b_{11}-b_2)}{\lambda^2}g(t_0)S^2.
\end{equation}
 In contrast, the linear extensional flow at $\rm{Wi \gtrsim 1}$ and $\rm{De \ll 1}$ satisfies a different relationship:
\begin{equation}\label{stretching_stress_relation_strong_flow_lin_ext}
{\rm tr}(\boldsymbol{\tau_p})=2\frac{(b_{11}-b_2)}{\lambda^2}g(t_0)(ln(S))^2.
\end{equation}
}
The stress field grows quadratically with Wi for both the shear and extensional flows (\textit{SI Appendix}). However, the stretching field at a large Wi grows linearly with Wi for shear flows, whereas it grows exponentially for linear extensional flow, leading to the slower growth of ${\rm tr}(\boldsymbol{\tau_p})$ with $S$ in the extensional flow than in the shear flows.

\textcolor{black}{
Thus, in ordered and MWCSH flows, ${\rm tr}(\boldsymbol{\tau_p})$ and $S$ are related by compact analytical expressions for different steady (Eq. \ref{stretching_stress_relation_general}) and weakly unsteady (Eq. \ref{stretching_stress_relation_unsteady}) flows at $\rm{Wi<1}$. 
Further, the analysis of flows undergoing MWCSH uncovers exact relationships between ${\rm tr}(\boldsymbol{\tau_p})$ and $S$ at $\rm{Wi \gg 1}$ for pure shear (Eq. \ref{stretching_stress_relation_strong_flow_shear}) and pure extensional (Eq. \ref{stretching_stress_relation_strong_flow_lin_ext}) flows (\textit{SI Appendix}). 
These results clearly illustrate a deep-seated quantitative relationship between the polymeric stress and Lagrangian stretching history, which links the topologies of these fields. 
However, in general for flows having mixed-kinematics, such exact expressions at $\rm{Wi>1}$ are not accessible. 
Hence, we use numerical simulations to further explore the relationship between the stress and stretching fields in complex geometries at large Wi.
}
 
\begin{figure*}[!ht]
 \centering
 \includegraphics[width=.95\textwidth]{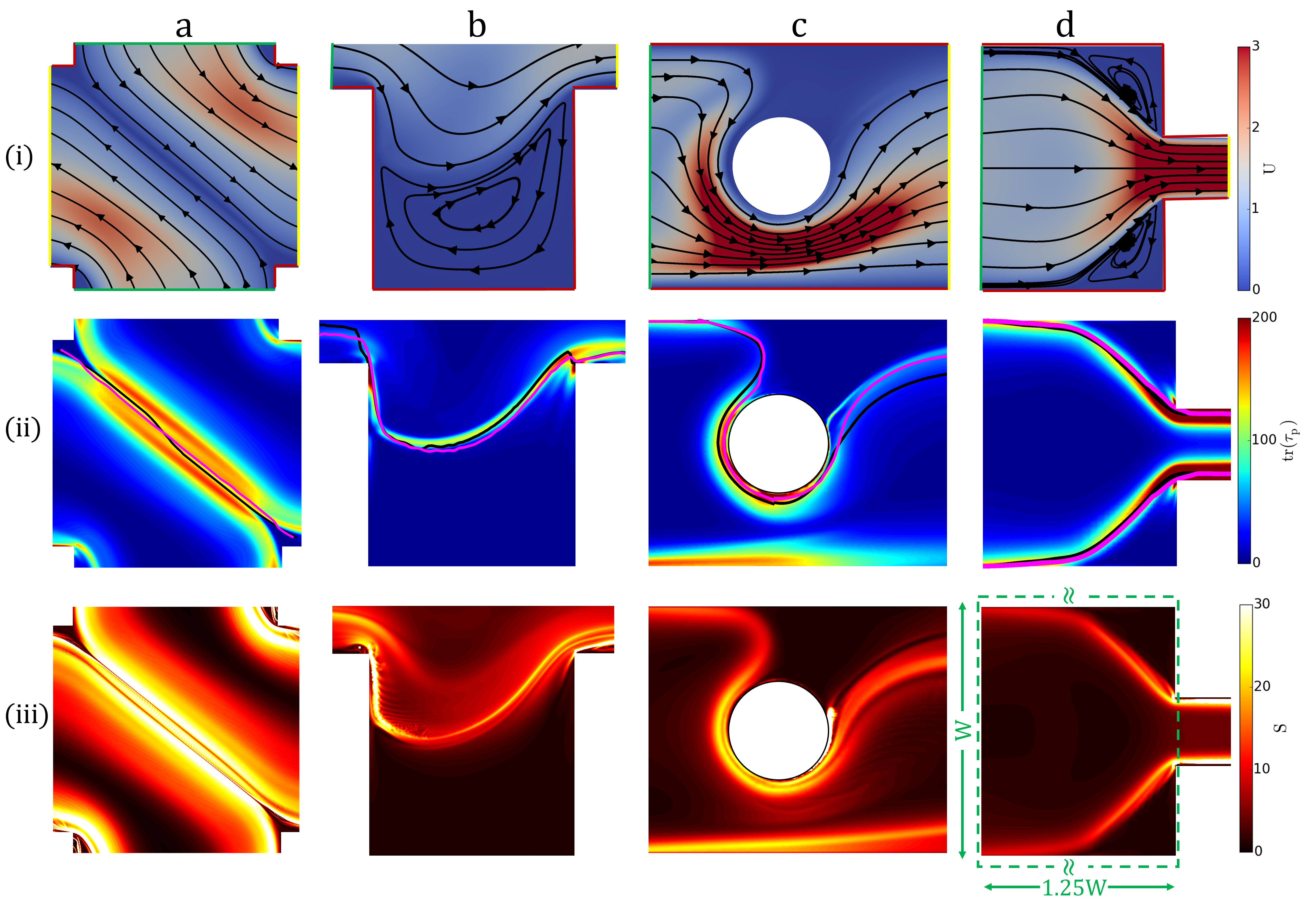}
 \caption{\label{isolated_geometries.png} Flow field (row i), trace of the polymeric stress tensor (row ii), and stretching field (row iii) for viscoelastic flows in different geometries at large Weissenberg number (${\rm Wi} \gtrsim 1$): (column a) cross-slot geometry at ${\rm Wi}=4$, (column b) flow over a cavity at ${\rm Wi}=1.25$, (column c) cylinder confined in a channel at ${\rm Wi}=2.5$, and (column d) flow through an isolated constriction at ${\rm Wi}=0.75$. 
 ${\rm Wi}=\lambda U_{in}/L_c$, where $U_{in}$ is the inlet velocity of the geometry. \textcolor{black}{The characteristic length scale, $L_c$, for the geometries are the upstream channel width (a,d), the channel width at the cavity (b), and the cylinder diameter (c).} $\rm{U}$ is normalized by $U_{in}$, and $\boldsymbol{\tau_p}$ is normalized by $\eta_0 U_{in}/L_c$, where $\eta_0$ is the zero-shear rate viscosity of the viscoelastic fluid. \textcolor{black}{Black and magenta lines (row ii) are the stretching manifolds (ridge of maximal stretching) obtained from stretching fields (row iii; \textcolor{black}{see also \textit{SI Appendix} Fig.~S5}) for integration time intervals of $\Delta t=\lambda$ and $\Delta t=2\lambda$, respectively. The stretching fields shown (row iii) correspond to $\Delta t=\lambda$.} 
 No-slip boundaries are highlighted by red lines, whereas inlets and outlets are indicated by green and yellow, respectively (row i). 
 Geometries shown are a small portion of larger simulation domains, which ensured sufficient entrance and exit lengths.
 }
 \end{figure*}

\subsection*{\textcolor{black}{Numerical simulations of stress and stretching for strongly viscoelastic flows}}
Beyond the exact analytical correspondence between stress and stretching for weakly unsteady viscoelastic flows (Eqs. \ref{stretching_stress_relation_unsteady}, \ref{stretching_stress_relation_strong_flow_shear}, and \ref{stretching_stress_relation_strong_flow_lin_ext}), strong nonlinearities yield complex and time-dependent flow structures that emerge at large Weissenberg number \cite{Pakdel1996,Kumar2021multistability,Kumar2021tristability}.
To illustrate the persistent concordance between the polymeric stress and stretching field topologies, viscoelastic flows are numerically simulated through various geometries at large Weissenberg numbers (${\rm Wi} \gtrsim 1$). 
The polymeric stress tensor and velocity field are calculated using a FENE-P constitutive model, which captures fluid elasticity and shear-thinning behaviors, as well as the finite stretching of the polymeric chains \cite{bird1987dynamics_vol2}:
\begin{equation}\label{fenep}
\boldsymbol{\tau}_p+\frac{\lambda}{f}\overset{\nabla}{\boldsymbol{\tau}}_p=\frac{a\eta_p}{f}(\nabla \bold{u}+\nabla \bold{u}^\intercal)-\frac{D}{Dt}\left(\frac{1}{f}\right)[\lambda \boldsymbol{\tau}_p+a\eta_p \bold{I}],
\end{equation}
where $\eta_p$ is the polymeric contribution to the zero-shear rate viscosity of the fluid.
$\overset{\nabla}{\boldsymbol{\tau}}_p$ is given by:
\begin{equation}\label{hat_delta}
\overset{\nabla}{\boldsymbol{\tau}}_p=\frac{D\boldsymbol{\tau}_p}{Dt}-\boldsymbol{\tau}_p\cdot\nabla\bold{u}-\nabla\bold{u}^\intercal \cdot\boldsymbol{\tau}_p,
\end{equation}
and the nonlinear function $f$ is: 
\begin{equation}\label{f_taup}
f(\boldsymbol{\tau_p})=\frac{L^2+\frac{\lambda}{a\eta_p} {\rm tr}(\boldsymbol{\tau_p})}{L^2-3},
\end{equation}
where $a=L^2/(L^2-3)$ and $L$ is the maximum extensibility of the polymeric chains. 
Numerical simulations are implemented using OpenFOAM \cite{Jasak2007} and RheoTool \cite{Pimenta2017}. 
The log-conformation method is used to solve for the logarithm of the conformation tensor ($\boldsymbol{\Theta}$) \cite{Fattal2004,Pimenta2017}, and then the polymeric stress tensor is determined using:
\begin{equation}\label{log_conf}
\boldsymbol{\tau}_p=\frac{\eta_p}{\lambda}(f e^{\boldsymbol{\Theta}}-a\bold{I}).
\end{equation}
The stretching field ($S$) is also calculated numerically from the simulated velocity field:
Four auxiliary points centered around each primary grid point define a fluid element. 
The flow map ($ \boldsymbol{\Phi}$) is obtained by numerically integrating the auxiliary point position in time, and the deformation-gradient tensor ($\nabla \boldsymbol{\Phi}$) on each primary grid point is computed by central differencing of the auxillary points~\cite{Onu2015}.

At large Weissenberg number, the Lagrangian stretching field mirrors the stress field topology across four different benchmark geometries (Fig. \ref{isolated_geometries.png}). 
For ${\rm Wi}>{\rm Wi}_{cr}$, strong flow asymmetries develop in the hyperbolic base flow of the cross-slot geometry (Fig. \ref{isolated_geometries.png}a(i)) \cite{Poole2007,Haward2016}, as well as in the flow past a confined cylinder (Fig. \ref{isolated_geometries.png}c(i)) \cite{Varchanis2020}. Despite the otherwise creeping flow conditions, viscoelasticity leads to flow separation in the corners upstream of an isolated constriction (Fig. \ref{isolated_geometries.png}d(i)) \cite{Lanzaro2011} as well as an unsteady asymmetric eddy in the flow over a cavity (Fig. \ref{isolated_geometries.png}b(i)).
For all four geometries, the stretching field (Fig.~\ref{isolated_geometries.png}(iii)) has a strong correlation with the topology of the stress field (Fig.~\ref{isolated_geometries.png}(ii); \textcolor{black}{see \textit{SI Appendix} Fig.~S6)}, which are characterized by thin streaks with high values of $S$ and ${\rm tr}(\boldsymbol{\tau_p})$, respectively. 
These features indicate regions where polymers have experienced significant deformation -- and thus, stress -- due to the integrated effects of shear and extensional flow over the past $\Delta t = \lambda$.
\textcolor{black}{Our observations persist in three-dimensional flows (\textit{SI Appendix}, Fig.~S7) and are independent of the rheological model (\textit{SI Appendix}, Fig.~S8).}
\textcolor{black}{
The (attractive or unstable) stretching manifolds were extracted from the ridges of the maximal stretching for different integration times ($\Delta t =\lambda, 2\lambda$; Fig.~\ref{isolated_geometries.png}(iii); \textit{SI Appendix}, Fig.~S5) and superimposed on the stress field (Fig.~\ref{isolated_geometries.png}(ii)). 
In line with their known behavior as strong transport barriers, these material lines act as separatrices between regions with disparate flow characteristics, including asymmetric flows (Fig.~\ref{isolated_geometries.png}a,c) and separated eddies (Fig.~\ref{isolated_geometries.png}b,d).
While the magnitude of stretching increases with the integration time (\textit{SI Appendix}, Fig.~S5), the position of the stretching manifolds exhibits minimal change for $\Delta t > \lambda$, and they remain coincident with streaks of the stress fields (Fig.~\ref{isolated_geometries.png}(row ii)). 
}

\begin{figure}[!ht]
 \centering
 \includegraphics[width=.8\textwidth]{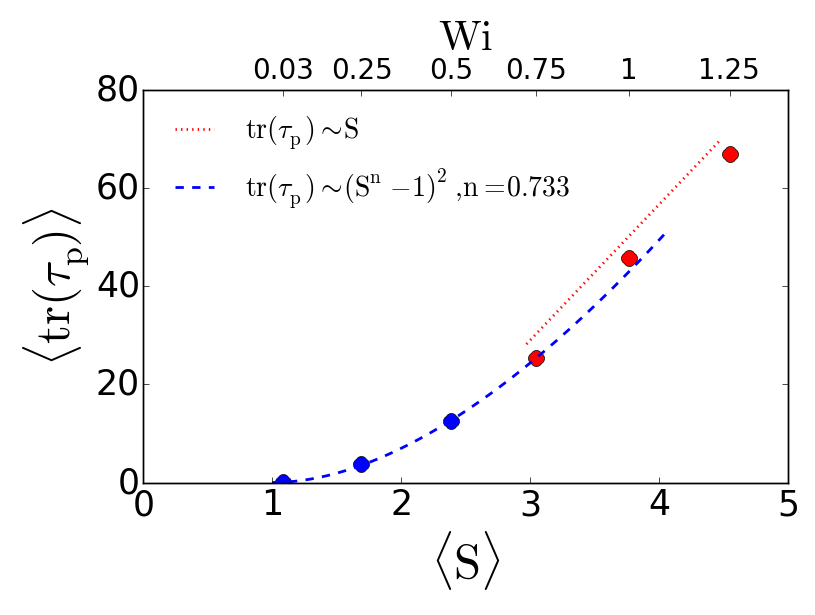}
 \caption{\label{trtaup_s_scaling.png} \textcolor{black}{Mean stress increases with mean stretching ($\Delta t=\lambda$) for viscoelastic flow through an isolated constriction (Fig.~\ref{isolated_geometries.png}d) at different $\rm{Wi}$. The region $\rm 1.25W \times W$ upstream of the constriction (Fig.~\ref{isolated_geometries.png}d(iii), green box), where $\rm W$ is the upstream width of the channel, is used to calculate the spatial average of stress and stretching. $\boldsymbol{\tau_p}$ is normalized by $\eta_0 U_{in}/L_c$ corresponding to $\rm Wi=0.75$.}}
 \end{figure}

\textcolor{black}{
Beyond comparing their respective topologies, numerical simulations of viscoelastic flows enable us to further investigate the quantitative relationship between the magnitude of the stress and stretching fields.
As an illustrative example, we consider the spatial average of the stress, $\langle {\rm tr(\tau_p)} \rangle$, and stretching, $\langle S \rangle$, over a fixed region of space (Fig.~\ref{trtaup_s_scaling.png}) within the constriction flow (Fig.~\ref{isolated_geometries.png}d(iii), green box).
For small Weissenberg number, the predicted scaling ${\rm tr(\tau_p)} \sim (S^n-1)^2 $ (Eq. \ref{stretching_stress_relation_general}) is recovered as $\rm{Wi \to 0}$ (Fig.~\ref{trtaup_s_scaling.png}, blue), similar to the scaling between the local values of stress and stretching fields (\textit{SI Appendix}, Fig.~S10). 
\textcolor{black}{
At large Wi, $\langle {\rm tr}(\boldsymbol{\tau_p}) \rangle$ and $\langle S \rangle$ exhibit a linear scaling (Fig. \ref{trtaup_s_scaling.png}, red), which we hypothesize is due to the highly-mixed flow kinematics upstream of the constriction (\textit{SI Appendix}, Fig.~S1):
In this regime, the slope of ${\rm tr}(\boldsymbol{\tau_p})-S$ curve increases with $S$ for purely shear deformation (Eq. \ref{stretching_stress_relation_strong_flow_shear}) but decreases for purely extensional deformation (i.e., for $S>e$) (Eq. \ref{stretching_stress_relation_strong_flow_lin_ext}). In this example, the quantitative relationship between the stress and stretching fields is obtained along with their topological resemblance. However, the scaling exponent is not universal (\textit{SI Appendix}, Fig.~S10), and developing a robust general predictive framework for relating ${\rm tr}(\boldsymbol{\tau_p})$ and $S$ will require further investigation.
}
} 

\begin{figure*}[!ht]
 \centering
 \includegraphics[width=\textwidth]{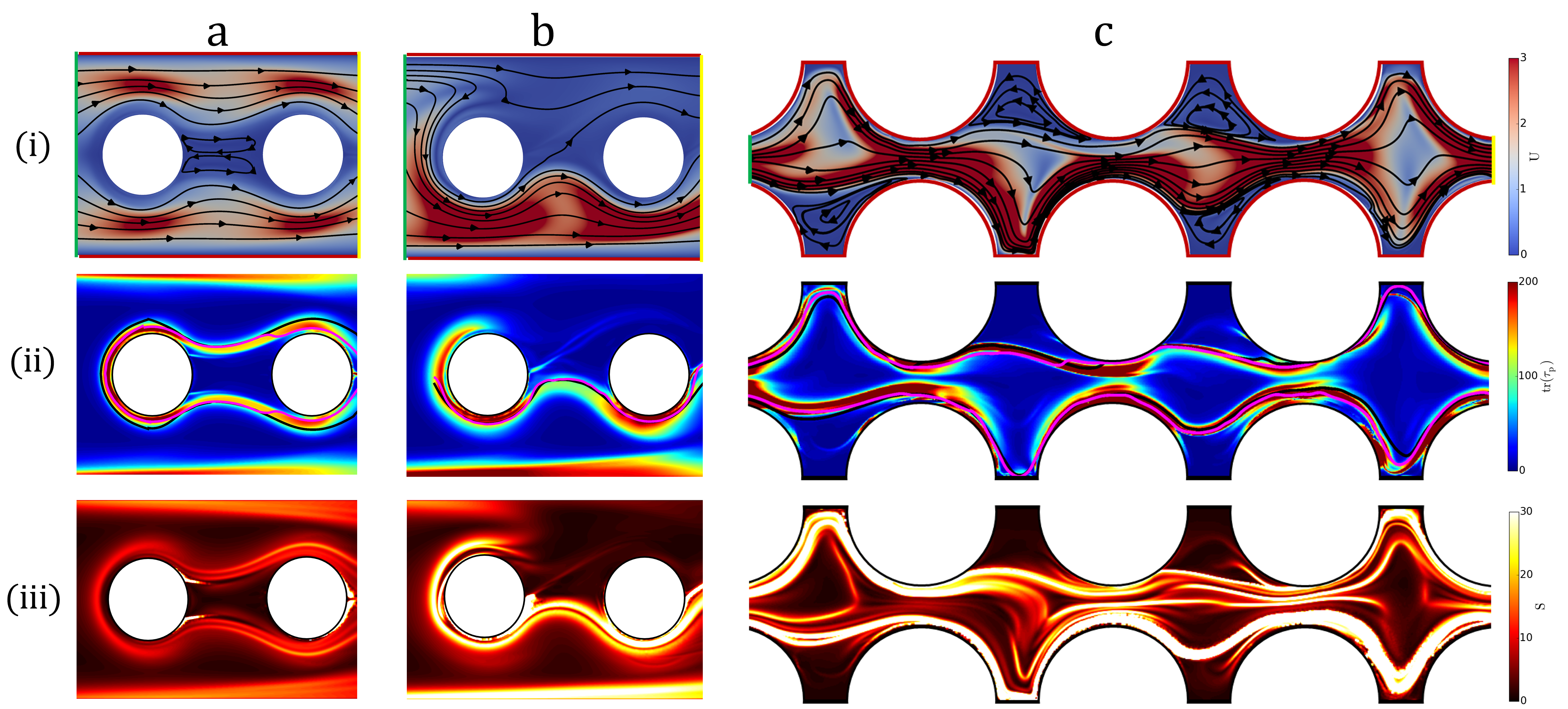}
 \caption{\label{viscoelastic_instability.png} \textcolor{black}{Instantaneous} flow fields (row i), trace of the polymeric stress tensor (row ii), and stretching field (row iii) for flow states, stemming from viscoelastic instabilities. 
 Two streamwise cylinders in a channel at (column a) moderate (${\rm Wi}=1.88$) and (column b) large (${\rm Wi}=3.12$) Weissenberg number.
 (column c) Corrugated channel at ${\rm Wi}=1.68$ \textcolor{black}{(see also \textit{SI Appendix}, Fig.~S9)}. 
 ${\rm Wi}=\lambda U_{in}/L_c$, where $L_c$ is the cylinder diameter in (a) and (b) and the half-width of the pore for (c). $U$ and $\boldsymbol{\tau_p}$ are normalized by $U_{in}$ and characteristic shear stress ($\eta_0 U_{in}/L_c$), respectively. \textcolor{black}{Black and magenta lines (row ii) represent the stretching manifolds (ridge of maximal stretching) obtained from stretching fields
(row iii; see also \textit{SI Appendix}, Fig.~S5)  for integration time intervals of $\Delta t=\lambda$ and $\Delta t=2\lambda$, respectively. 
 The stretching fields shown (row iii) correspond to $\Delta t=\lambda$.}
 No-slip boundaries are highlighted by red lines, whereas inlets and outlets are green and yellow, respectively (row i).
 Geometries shown are a small portion of larger simulation domains, which ensured sufficient entrance and exit lengths.
 }
 \end{figure*}

\subsection*{\textcolor{black}{Extension to complex geometries}}
Intricate flow geometries -- for example in porous media~\cite{Walkama2020} -- increase the complexity of instabilities leading to multi-stable and strongly time-dependent flow structures~\cite{Kumar2021tristability,Kumar2021multistability,Varshney2017}.
Here, we compare the polymeric stress field and Lagrangian stretching field for topologically complex and unsteady flows stemming from multiple cylinders \cite{Varshney2017,Kumar2021tristability} and constrictions \cite{Kumar2021multistability}. 
The addition of a second cylinder in a steady channel flow leads to two transitions with increasing ${\rm Wi}$~\cite{Varshney2017,Kumar2021tristability}: 
At the first transition, the elastic wake downstream of the first cylinder bifurcates, yielding two symmetric eddies (Fig. \ref{viscoelastic_instability.png}a(i)) encircled by streaks of high stress (Fig. \ref{viscoelastic_instability.png}a(ii)).
At the second transition, the stress topology becomes asymmetric (Fig.~\ref{viscoelastic_instability.png}b(ii)), and the flow is diverted to one side of the cylinders (Fig. \ref{viscoelastic_instability.png}b(i)).
Subsequent to each transition, ridges of $S$ coincide with regions of high ${\rm tr}(\boldsymbol{\tau_p})$ (Fig. \ref{viscoelastic_instability.png}a(ii) and \ref{viscoelastic_instability.png}b(ii)), whereby the stretching manifolds isolate the regions of vortical and quiescent flow from the bulk.
Finally, viscoelastic instability in flow through a series of interconnected pores and throats leads to fluctuating flow patterns (Fig. \ref{viscoelastic_instability.png}c(i); \textit{SI Appendix}, Fig.~S9) \cite{Kumar2021multistability}. 
Flow separation in the high-stress throat (Fig. \ref{viscoelastic_instability.png}c(ii)) causes eddy formation in different regions of the pores, corresponding to four distinct flow patterns (Fig. \ref{viscoelastic_instability.png}c(i)).
For large Wi, advection of the stressed polymers spans multiple pores, and consequently, the Lagrangian stretching field exhibits a richer topology (Fig. \ref{viscoelastic_instability.png}c(iii)).
However, the LCSs expected to dominate the dynamics are the strongest local stretching lines~\cite{Haller2015}, which indeed correspond to ridges in the stress field (Fig. \ref{viscoelastic_instability.png}c(ii)). 
\textcolor{black}{
The secondary ridges and finer structures in the stretching field of unsteady flows (Fig. \ref{viscoelastic_instability.png}c(iii)) emerge due to mixed kinematics (\textit{SI Appendix}, Fig.~S2), and they are further refined with increasing integration time ($\Delta t$; \textit{SI Appendix}, Fig.~S5). 
However, the maximum attractive (unstable) material lines (strongest ridges), which control the flow states, remain nearly unchanged (Fig. \ref{viscoelastic_instability.png}c(ii)).
}

\textcolor{black}{
At large Wi, despite the excellent agreement between the stress field and stretching field in the regions of high stress, subtle differences also persist in other regions (Figs.~\ref{viscoelastic_instability.png}c(ii) and c(iii)). 
The regions where streaks of high polymeric stress form are largely shear-dominated (\textit{SI Appendix}, Figs.~S1~and~S2), as they act as barriers to flow crossing and exist between regions of strong extensional or vortical flow~\cite{Kumar2021multistability}. 
The theoretic analysis of MWCSH flows has shown that there exits a direct relationship between the stress and stretching for both homogeneous and non-homogeneous shear flows at $\rm{Wi \gg 1}$ (Eq. \ref{stretching_stress_relation_strong_flow_shear}). 
Taken together, these results illustrate the origin of the strong correlation between the stress and stretching fields in the regions of high stress.
In contrast, mixed-kinematics regions away from high stress zones (\textit{SI Appendix}, Figs.~S1~and~S2) include extensional flow components with a different stress-stretching scaling at $\rm{Wi>1}$ (Eqs. \ref{stretching_stress_relation_strong_flow_shear} and \ref{stretching_stress_relation_strong_flow_lin_ext}).
\textcolor{black}{Detailed numerical analysis supports this observation and indicates decreased correlation between the stress and stretching fields as mixed kinematics emerge in strong flows (${\rm Wi} \gtrsim O(1)$; \textit{SI Appendix}, Figs.~S3 and S4).}
}

\section*{Conclusions}
Knowledge of the stress field is essential to elucidate the emergent flow patterns and transport properties in viscoelastic flows. 
The work presented here applies concepts from Lagrangian coherent structures to gain new insights into the often Eulerian framework of viscoelastic fluid mechanics, thus bringing together two disparate fields of continuum analysis.
\textcolor{black}{In doing so, we show that the stretching field, which only depends on the flow kinematics, is a powerful indicator of the underlying polymeric stress field.}
For small Wi, we analytically derived a general relationship between the trace of polymeric stress tensor and the Lagrangian stretching field, and for unstable flows at large Wi, numerical simulations show a strong correlation between the stress topology and manifolds of the stretching field. The extension of these results to three dimensions provides copious opportunities for future investigations.
An important outcome of this work is the potential to determine the stress field topology directly from conventional experimental velocimetry data for arbitrary viscoelastic materials and flow geometries.
LCSs that underlie turbulent and chaotic flows are known to regulate material transport, and anomalous transport effects often arise from unstable polymeric flows~\cite{Walkama2020, Yamani2021}. The concepts established here show intriguing links between polymeric stress and stretching kinematics, which could prove useful in investigating the dynamics and transport for a range of applications from mixing to natural flows~\cite{Walkama2020,Groisman2001,Tang2013,Suarez2006}.

\section*{Acknowledgement}
We thank D.M. Walkama for helpful discussions. 
This work was supported by National Science Foundation awards CBET-1700961, CBET-1705371, and CBET-2141404 (to A.M.A.), and CBET-2141349, CBET-1701392, and CAREER-1554095 (to J.S.G.).


\section*{Appendix}
\renewcommand\thefigure{S\arabic{figure}}
\setcounter{figure}{0}

\subsection*{Weakly unsteady viscoelastic flows }
The ordered fluid model can be applied to weakly unsteady viscoelastic flows at small Weissenberg number, $\rm{Wi \ll 1}$. Here, we introduce a weak unsteadiness in the analytical flows studied in the main text as $\boldsymbol{u}=\boldsymbol{u_0}[1+ {\rm De} \alpha(t)]$, where $\boldsymbol{u_0}$ is the steady linear or nonlinear flow, $\rm De$ is the Deborah number, and $\alpha(t)$ is an arbitrary time-dependent function. In the limit of weak viscoelasticity ($\rm{Wi \ll 1}$) and weak unsteadiness ($\rm{De \ll 1}$), we derive both the stress field and the stretching field at time $t=t_0$. For linear unsteady extensional flow, the velocity components are $u=\dot{\epsilon} x\{1+{\rm De} \alpha(t)\}$ and $v=-\dot{\epsilon} y\{1+{\rm De} \alpha(t)\}$, which lead to the following stress and stretching fields:

\begin{equation}\label{tau_linear_ext_si}
{\rm tr}(\boldsymbol{\tau_p})=\frac{8(b_{11}-b_2)}{\lambda^2}\{1+{\rm De} \alpha(t_0)\}^2{\rm Wi}^2,
\end{equation}

\begin{equation}\label{S_linear_ext_si}
S^2=e^{2{\rm Wi}\{1+{\rm De} \beta(t_0)\}}=1+2\{1+{\rm De} \beta(t_0)\}{\rm Wi}+2\{1+{\rm De} \beta(t_0)\}^2{\rm Wi}^2+H.O.T.,
\end{equation}
where ${\rm Wi=}\dot{\epsilon}\lambda$ and $\beta(t_0)=-\frac{1}{\lambda}\int_{t_0}^{t_0-\lambda} \alpha (t) dt$. For simple shear flow ($u=\dot{\gamma} y\{1+{\rm De} \alpha(t)\}, v=0$), the stress and stretching fields are:
\begin{equation}\label{tau_linear_shear_si}
{\rm tr}(\boldsymbol{\tau_p})=\frac{2(b_{11}-b_2)}{\lambda^2}\{1+{\rm De} \alpha(t_0)\}^2{\rm Wi}^2,
\end{equation}

\begin{align}\label{S_linear_shear_si}
S^2&=1+\frac{1}{2}\{1+{\rm De} \beta(t_0)\}^2{\rm Wi}^2+\{1+{\rm De} \beta(t_0)\}{\rm Wi} \left(1+\frac{1}{4}\{1+{\rm De} \beta(t_0)\}^2{\rm Wi}^2 \right)^{1/2} \\& =1+\{1+{\rm De} \beta(t_0)\}{\rm Wi}+\frac{1}{2}\{1+{\rm De} \beta(t_0)\}^2{\rm Wi}^2+H.O.T.,
\end{align}
where $\rm{Wi=}\dot{\gamma}\lambda$. For rotational flow ($u=-\Omega y\{1+{\rm De} \alpha(t_0)\}$,~~$v=\Omega x\{1+{\rm De} \alpha(t_0)\}$), the stress and stretching are:
\begin{equation}\label{tau_linear_rot_si}
{\rm tr}(\boldsymbol{\tau_p})=0,
\end{equation}
\begin{equation}\label{S_linear_rot_si}
S^2=1.
\end{equation}

We also obtain stress and stretching for non-linear flows. In a weakly unsteady quadratic extensional flow with $u=\dot{\epsilon_1} xy\{1+{\rm De} \alpha(t)\}$ and  $v=-\frac{1}{2}\dot{\epsilon_1} y^2\{1+{\rm De} \alpha(t)\}$, the stress and stretching fields are given by:
\begin{equation}\label{tau_ext_quad_si}
{\rm tr}(\boldsymbol{\tau_p})=\frac{2(b_{11}-b_2)}{\lambda^2}(\Bar{x}^2+4\Bar{y}^2)\{1+{\rm De} \alpha(t_0)\}^2{\rm Wi}^2,
\end{equation}
\begin{align}\label{s_ext_quad_si}
S^2=1&+\left(\Bar{x}^2+4\Bar{y}^2\right)^{1/2}\{1+{\rm De} \beta(t_0)\}{\rm Wi} \\ &+\frac{1}{2}\left\{\left(\Bar{x}^2+4\Bar{y}^2\right)+\left( \frac{2\Bar{y}^3-\Bar{y}\Bar{x}^2}{\sqrt{\Bar{x}^2+4\Bar{y}^2}}\right)\right\}\{1+{\rm De} \beta(t_0)\}^2{\rm Wi}^2+H.O.T.,
\end{align}
where ${\rm Wi}=\dot{\epsilon_1}L_c\lambda$, $\Bar{x}=x/L_c$, and $\Bar{y}=y/L_c$. $L_c$ is the characteristic length scale of the flow. For a Poiseuille flow through a channel of height $2H$ with center-line flow speed $U_0$, the velocity field components are $u=U_0[1-(y/H)^2]\{1+{\rm De} \alpha(t)\}$ and $v=0$.
The resulting stress and stretching fields are:
\begin{equation}\label{tau_pois_si}
{\rm tr}(\boldsymbol{\tau_p})=\frac{8(b_{11}-b_2)}{\lambda^2}\Bar{y}^2\{1+{\rm De} \alpha(t_0)\}^2{\rm Wi}^2, 
\end{equation}
\begin{align}\label{s_pois_si}
S^2 &=1+  2\Bar{y}^2\{1+{\rm De} \beta(t_0)\}^2{\rm Wi}^2+ 2\Bar{y}\{1+{\rm De} \beta(t_0)\} {\rm Wi} \left(1+\Bar{y}^2\{1+{\rm De} \beta(t_0)\}^2{\rm Wi}^2 \right)^{1/2} \\ 
&= 1+2\Bar{y}\{1+{\rm De} \beta(t_0)\}{\rm Wi}+2\Bar{y}^2\{1+{\rm De} \beta(t_0)\}^2{\rm Wi}^2+H.O.T.,
\end{align}
where ${\rm Wi}=U_0\lambda/H$ and $\Bar{y}=\lvert y \rvert/H$. For the quartic extensional flow, the velocity field components are $u=\frac{1}{2}\dot{\epsilon_2} x^2y^2\{1+{\rm De} \alpha(t)\}$ and $v=-\frac{1}{3}\dot{\epsilon_2} xy^3\{1+{\rm De} \alpha(t)\}$, and analytically derived stress and stretching fields are:
\begin{equation}\label{tau_quart_si}
{\rm tr}(\boldsymbol{\tau_p})=\frac{2(b_{11}-b_2)}{\lambda^2}\left(\Bar{x}^4\Bar{y}^2+\frac{10}{3}\Bar{x}^2\Bar{y}^4+\frac{1}{9}\Bar{y}^6\right)\{1+{\rm De} \alpha(t_0)\}^2{\rm Wi}^2, 
\end{equation}
\begin{align}\label{s_ext_quart_si}
S^2=1 &+ \left(\Bar{x}^4\Bar{y}^2+\frac{10}{3}\Bar{x}^2\Bar{y}^4+\frac{1}{9}\Bar{y}^6\right)^{1/2}\{1+{\rm De} \beta(t_0)\}{\rm Wi} \\ & +\frac{1}{2}\left\{\left(\Bar{x}^4\Bar{y}^2+\frac{10}{3}\Bar{x}^2\Bar{y}^4+\frac{1}{9}\Bar{y}^6\right)+\frac{1}{3}\left( \frac{-6\Bar{x}^5\Bar{y}^4+5\Bar{x}^3\Bar{y}^6+\Bar{x}\Bar{y}^8}{\sqrt{9\Bar{x}^4\Bar{y}^2+30\Bar{x}^2\Bar{y}^4+\Bar{y}^6}}\right)\right\}\{1+{\rm De} \beta(t_0)\}^2 {\rm Wi}^2 \\ & +H.O.T.,
\end{align}
where ${\rm Wi}=\dot{\epsilon_2}L_c^3\lambda$, $\Bar{x}=x/L_c$, and $\Bar{y}=y/L_c$. $L_c$ is the length scale of the flow. For all these weakly unsteady viscoelastic flows, the stress field (${\rm tr}(\boldsymbol{\tau_p})$) and stretching field ($S$) satisfy the following relationship:
\begin{equation}\label{stretching_stress_relation_unsteady_si}
{\rm tr}(\boldsymbol{\tau_p})= \frac{8}{n^2}\frac{(b_{11}-b_2)}{\lambda^2}g(t_0)(S^n-1)^2,
\end{equation}
where
\begin{equation}\label{order_unsteady_g_si}
g(t_0)=\left [\frac{1+{\rm De}\alpha(t_0)}{1+{\rm De}\beta(t_0)} \right ]^2,
\end{equation}
$n\neq 0$, and $\beta(t_0)=-\frac{1}{\lambda}\int_{t_0}^{t_0-\lambda} \alpha (t) dt$. The polymeric relaxation time for the second-order fluid model can be given as $\lambda =-b_2/b_1$ \cite{bird1987dynamics_vol1}. Hence, the prefactor of Eq. \ref{stretching_stress_relation_unsteady} can be further simplified as:
\begin{equation}\label{stretching_stress_relation_unsteady_2nd_prefactor}
{\rm tr}(\boldsymbol{\tau_p})= \frac{8}{n^2}\frac{b_1^2(b_{11}-b_2)}{b_2^2}g_1(t_0)(S^n-1)^2.
\end{equation}

\section*{Mapping of different viscoelastic models to the second-order fluid model}
\textcolor{black}{Different models of viscoelastic fluids converge to the second-order fluid model for $\rm Wi \ll 1$. The mapping of the model parameters is given in Table \ref{parameter_mapping}.}
 \begin{table}[h!]
 \caption{ \label{parameter_mapping} \textcolor{black}{The mapping of the parameters of different viscoelastic models to the second-order fluid model at $\rm Wi \ll 1$. $\eta_p$ and $\lambda$ are the polymeric contribution to the zero-shear rate viscosity and polymeric relaxation time, respectively. $\alpha$ is the mobility factor in the Giesekus model. Model parameters $L \to \infty$ and $\epsilon \to 0$ for the second-order fluid expansion of FENE-P and sPTT models, respectively.}}
 \begin{center}
 \begin{tabular}{ |c|c|c|c| } 
 \hline
 \textbf{Second Order} & \textbf{FENE-P} & \textbf{Giesekus} & \textbf{sPTT}  \\
 \hline
 $b_1$ & $\eta_p$ & $\eta_p$ & $\eta_p$  \\
 \hline
 
  $b_2$ & $-\eta_p \lambda$ & $-\eta_p \lambda$ & $-\eta_p \lambda$  \\
 \hline
 
  $b_{11}$ & $0$ & $-\alpha\eta_p\lambda$ & $0$  \\
 \hline
 \end{tabular}
\end{center}
\end{table}

\subsection*{Motions with constant stretch history}
Linear and Poiseuille flows are characterized as \textit{motions with constant stretch history} (MWCSH) \cite{Noll1962,Huilgol1969,Huilgol1976,HUILGOL1971}. The stress tensor of such flows undergoing the MWCSH is:
\begin{equation}\label{tau_p_si}
\boldsymbol{\tau_p}=b_1 \boldsymbol{\gamma_{(1)}}+b_2 \boldsymbol{\gamma_{(2)}}+b_{11} \{\boldsymbol{\gamma_{(1)}} \cdot \boldsymbol{\gamma_{(1)}}\}+b_3 \boldsymbol{\gamma_{(3)}}+b_{12} \{\boldsymbol{\gamma_{(1)}} \cdot \boldsymbol{\gamma_{(2)}}+\boldsymbol{\gamma_{(2)}} \cdot \boldsymbol{\gamma_{(1)}}\}+b_{1:11} \{\boldsymbol{\gamma_{(1)}} \colon \boldsymbol{\gamma_{(1)}}\}\boldsymbol{\gamma_{(1)}},
\end{equation}
where $b_1$, $b_2$, $b_3$, $b_{11}$, $b_{12}$ and $b_{1:11}$ are constants.  We consider weakly unsteady flows having velocity field $\boldsymbol{u}=\boldsymbol{u_0}[1+ {\rm De} \alpha(t)]$, where $\boldsymbol{u_0}$ represents the steady base flow. For linear extensional, simple shear, rotational and Poiseuille flows, the traces of their respective stress tensors (${\rm tr}(\boldsymbol{\tau_p})$) at time $t=t_0$ are:

\begin{equation}\label{tau_linear_ext_csh_si}
{\rm tr}(\boldsymbol{\tau_p})=\left[\frac{8(b_{11}-b_2)}{\lambda^2}\{1+{\rm De} \alpha(t_0)\}^2+\frac{8(2b_{12}-3b_{3})}{\lambda^2}\{1+{\rm De} \alpha(t_0)\}{\rm De}\alpha^{\prime}(t_0)\right]{\rm Wi}^2,
\end{equation}

\begin{equation}\label{tau_linear_shear_csh_si}
{\rm tr}(\boldsymbol{\tau_p})=\left[\frac{2(b_{11}-b_2)}{\lambda^2}\{1+{\rm De} \alpha(t_0)\}^2+\frac{2(2b_{12}-3b_{3})}{\lambda^2}\{1+{\rm De} \alpha(t_0)\}{\rm De}\alpha^{\prime}(t_0)\right]{\rm Wi}^2,
\end{equation}

\begin{equation}\label{tau_linear_rotational_csh_si}
{\rm tr}(\boldsymbol{\tau_p})=0,
\end{equation}
and  
\begin{equation}\label{tau_Poi_csh_si}
{\rm tr}(\boldsymbol{\tau_p})=\left[\frac{8(b_{11}-b_2)}{\lambda^2}\{1+{\rm De} \alpha(t_0)\}^2+\frac{8(2b_{12}-3b_{3})}{\lambda^2}\{1+{\rm De} \alpha(t_0)\} {\rm De} \alpha^{\prime}(t_0)\right] \Bar{y}^2{\rm Wi}^2,
\end{equation}
respectively. 
For all these MWCSH flows (except rotational), ${\rm tr}(\boldsymbol{\tau_p})$ has only the $O(\rm{Wi^2})$ term. However, the stretching fields have multiple higher order terms. For $\rm{Wi<1}$, where the leading order term (i.e., $O({\rm Wi})$) of the stretching field still dominates, the relationship between ${\rm tr}(\boldsymbol{\tau_p})$ and $S$ for weakly unsteady MWCSH flows is:
\begin{equation}\label{stretching_stress_relation_mwcsh_si}
{\rm tr}(\boldsymbol{\tau_p})= \frac{8}{n^2}\frac{(b_{11}-b_2)}{\lambda^2}g(t_0)(S^n-1)^2,
\end{equation}
where
\begin{equation}\label{mwcsh_unsteady_g_si}
g(t_0)=\frac{\left[\{1+ {\rm De} \alpha(t_0)\}^2+\frac{2b_{12}-3b_3}{\lambda(b_{11}-b_2)}\{1+ {\rm De} \alpha(t_0)\} {\rm De} \zeta(t_0)\right]}{\left[1+ {\rm De} \beta(t_0)\right]^2},
\end{equation}
and $\zeta(t_0)=\lambda \alpha^{\prime}(t_0)$.  The $O({\rm Wi}^2)$ term of ${\rm tr}(\boldsymbol{\tau_p})$ has third order contributions ($b_3$ and $b_{12}$) due to the unsteadiness of the flow fields. Therefore, $g(t_0)$ also has coefficients associated with third order terms ($b_3$ and $b_{12}$).

The stress fields obtained for the flows undergoing MWCSH are also valid at $\rm{Wi \gtrsim 1}$. For strong kinematics ($\rm{Wi \gg 1}$), the $O(\rm{Wi^2})$ term of the stretching field dominates in the simple shear and Poiseuille flows, which leads to the following relationship between the stress and stretching fields at $\rm{Wi \gg 1}$ for both weakly unsteady ($\rm{De \ll 1}$) homogeneous and non-homogeneous shear flows: 
\begin{equation}\label{stretching_stress_relation_strong_flow_shear_si}
{\rm tr}(\boldsymbol{\tau_p})=2\frac{(b_{11}-b_2)}{\lambda^2}g(t_0)S^2.
\end{equation}
 However, the linear extensional flow at $\rm{Wi \gtrsim 1}$ and $\rm{De \ll 1}$ satisfies a different relationship:
\begin{equation}\label{stretching_stress_relation_strong_flow_lin_ext_si}
{\rm tr}(\boldsymbol{\tau_p})=2\frac{(b_{11}-b_2)}{\lambda^2}g(t_0)(ln(S))^2.
\end{equation}

\subsection*{Flow-type parameter}

We use the flow-type parameter ($\Lambda$) to characterize the local fluid deformation in mixed flows:
\begin{equation}\label{flow_type}
\rm{\Lambda=\frac{\lvert\mathbf{\dot{\gamma}} \rvert-\lvert\mathbf{\Omega}\rvert}{\lvert\mathbf{\dot{\gamma}}\rvert+\lvert\mathbf{\Omega}\rvert}},
\end{equation}
where $\rm{\mathbf{\dot{\gamma}}=\frac{1}{2}(\nabla \mathbf{u}+ \nabla \mathbf{u}^\intercal)}$ and $\rm{\mathbf{\Omega}=\frac{1}{2}(\nabla \mathbf{u} - \nabla \mathbf{u}^\intercal)}$ are the strain rate tensor and the vorticity tensor, respectively \cite{Kumar2022review}. The magnitude of the strain rate ($\lvert\mathbf{\dot{\gamma}} \rvert$) and vorticity ($\lvert\mathbf{\Omega} \rvert$) tensors are defined as $\lvert\mathbf{\dot{\gamma}} \rvert=\sqrt{2 \mathbf{\dot{\gamma}} \colon \mathbf{\dot{\gamma}}}$ and $\lvert\mathbf{\Omega} \rvert=\sqrt{2 \mathbf{\Omega} \colon \mathbf{\Omega}}$, respectively. The value of the flow type parameter ranges from $\rm{\Lambda=-1 }$ for pure rotational flow to $\rm{\Lambda=0}$ for pure shear flow to $\rm{\Lambda=1 }$ for pure extensional flow. The Poiseuille flow has spatially non-homogeneous deformation. However, the value of the flow-type parameter is $\Lambda=0$, indicating purely shear deformation. The flow-type for the quadratic extensional flow is
\begin{equation}\label{flow_type_quad}
{\rm \Lambda}=\frac{\sqrt{x^2+4y^2}-\lvert x \rvert}{\sqrt{x^2+4y^2}+\lvert x \rvert},
\end{equation}
and for the quartic extensional flow the flow-type is 
\begin{equation}\label{flow_type_quart}
{\rm \Lambda}=\frac{\sqrt{\frac{8x^2y^4}{3}+(x^2y+\frac{y^3}{3})^2}-\lvert x^2y+\frac{y^3}{3} \rvert}{\sqrt{\frac{8x^2y^4}{3}+(x^2y+\frac{y^3}{3})^2}+\lvert x^2y+\frac{y^3}{3} \rvert}.
\end{equation}
Hence, the non-linear extensional flows have spatially non-uniform flow-type ($\rm{0\leq \Lambda \leq 1}$) and exhibit mixed-kinematics. 

\begin{figure}[!ht]
 \centering
 \includegraphics[width=\textwidth]{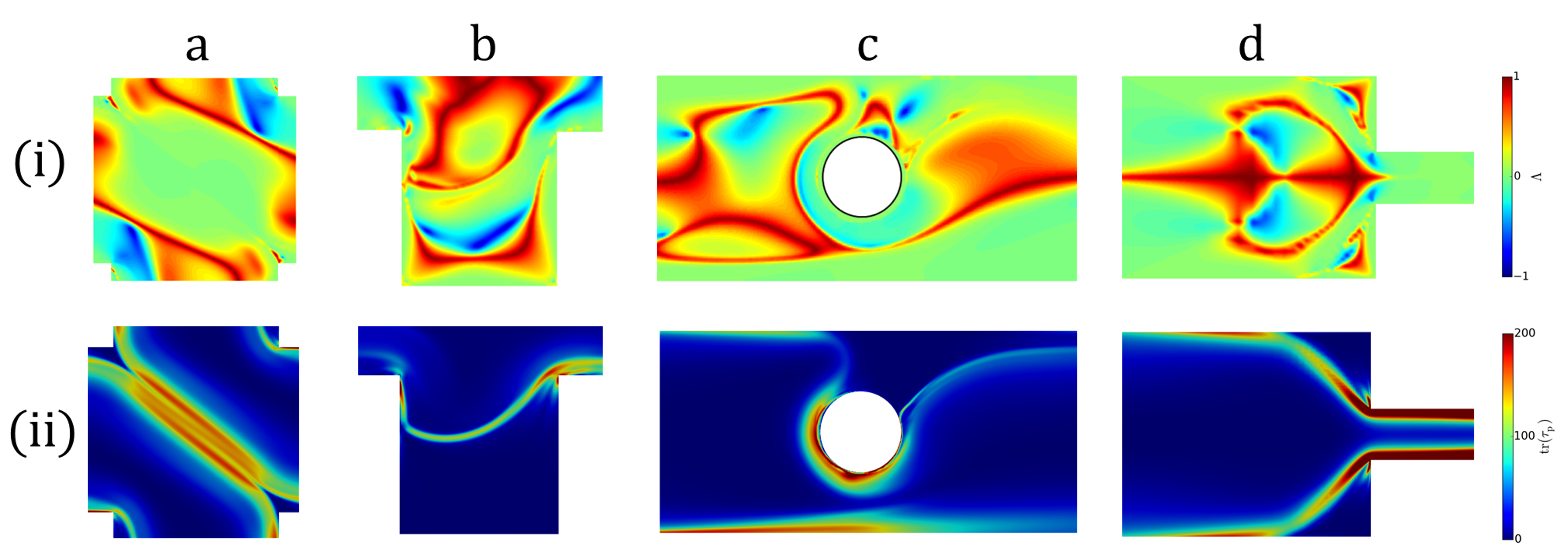}
 \caption{\label{flow_type_simple_geometry.png} (row ii) Flow-type parameter ($\Lambda$) and (row ii) the trace of polymeric stress tensor (${\rm tr}(\boldsymbol{\tau_p})$; repeated from Fig.~1 in the main text) for viscoelastic flows in benchmark geometries at large Weissenberg number (${\rm Wi} \gtrsim 1$): (column a) cross-slot geometry at ${\rm Wi}=4$, (column b) flow over a cavity at ${\rm Wi}=1.25$, (column c) cylinder confined in a channel at ${\rm Wi}=2.5$, and (column d) flow through an isolated constriction at ${\rm Wi}=0.75$. }
 \end{figure}

\begin{figure}[!ht]
 \centering
 \includegraphics[width=\textwidth]{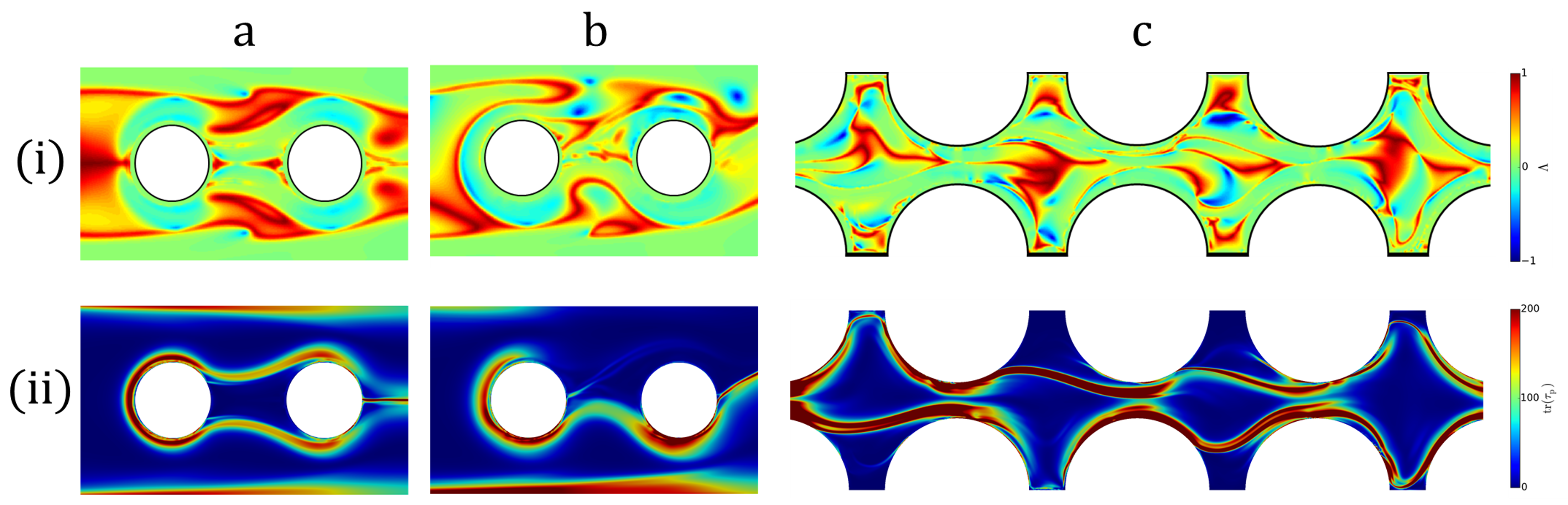}
 \caption{\label{flow_type_complex_geometry.png} (row ii) Flow-type parameter ($\Lambda$) and (row i) the trace of polymeric stress tensor (${\rm tr}(\boldsymbol{\tau_p})$; repeated from Fig.~3 in the main text) for viscoelastic flows in complex geometries at large Weissenberg number (${\rm Wi} \gtrsim 1$): (column a) two streamwise cylinders in a channel at moderate Wi (${\rm Wi}=1.88$),  (column b) two streamwise cylinders in a channel at large Wi (${\rm Wi}=3.12$), and
 (column c) corrugated channel at ${\rm Wi}=1.68$. }
 \end{figure}

The values of the flow-type parameter in benchmark and complex geometries investigated in the main text are shown in Fig.~\ref{flow_type_simple_geometry.png}(i) and Fig.~\ref{flow_type_complex_geometry.png}(i), respectively. For comparison with the flow-type, we also plot the polymeric stress field for the respective geometries (Figs. \ref{flow_type_simple_geometry.png}(ii) and \ref{flow_type_complex_geometry.png}(ii); repeated from main text Figs.~1~and~3). The formation of streaks characterized by large polymeric stress occurs in viscoelastic flows \cite{Kumar2021multistability} and these streaks act as barriers to flow crossing. Therefore, the regions where the streaks of large polymeric stress form are shear-dominated. The regions away from the streaks of high polymeric stress are predominantly extension-dominated or exhibit mixed-kinematics.     

\subsection*{Correlation between stress and stretching decreases in strong mixed kinematic flows} 
\begin{figure}[!ht]
 \centering
 \includegraphics[width=\textwidth]{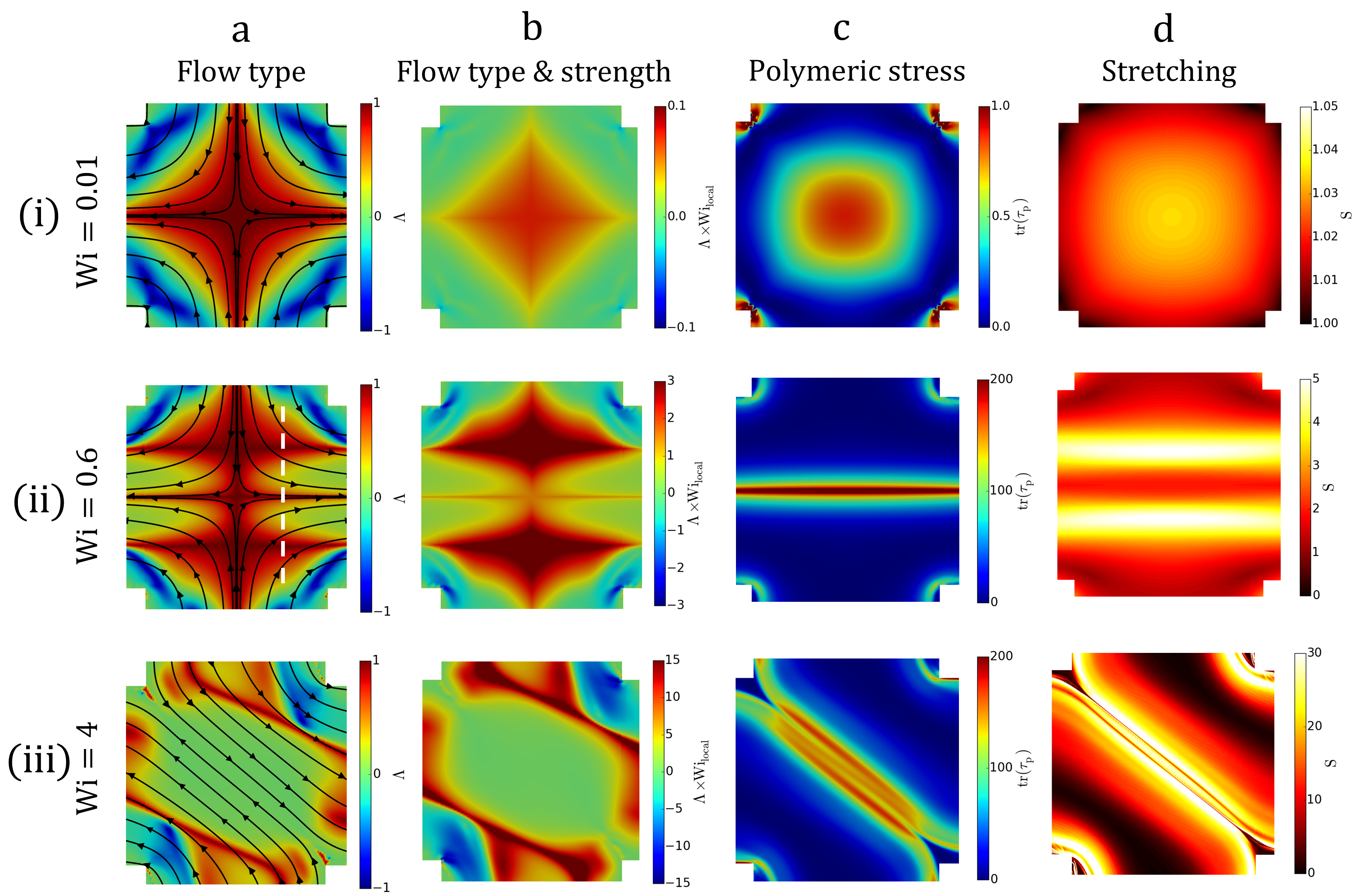}
 \caption{\label{cross_slot_strong_strain_rate.png}  (column a) Flow-type parameter, (column b) flow-type and strength, (column c) polymeric stress field, and (column d) stretching field in a viscoelastic flow through a cross-slot geometry at (row i) $\rm{Wi=0.01}$, (row ii) $\rm{Wi=0.6}$, and (row iii) $\rm{Wi=4}$.}
 \end{figure}

\begin{figure}[!ht]
\centering
\begin{subfigure}[b]{0.49\textwidth} 
\includegraphics[width=\textwidth]{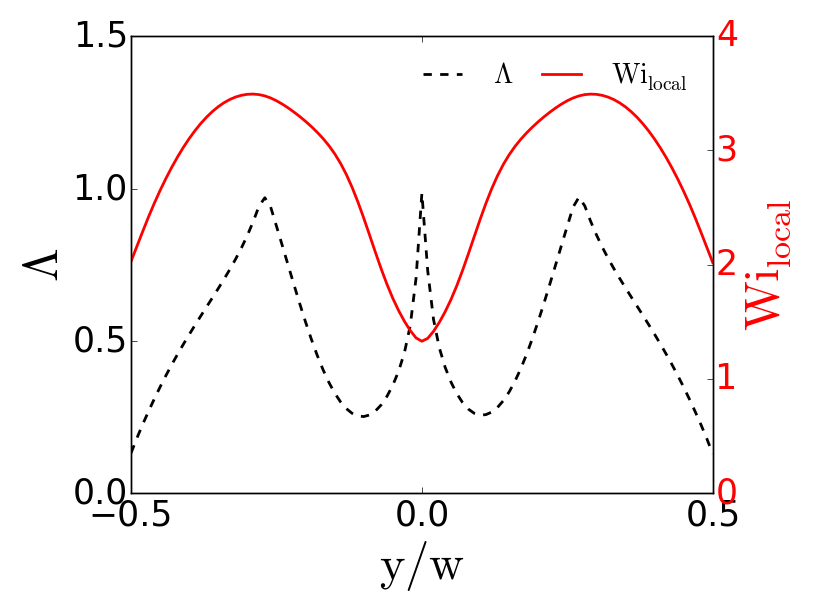}
\caption{}
\label{flow_type_Wi_local_u600ums_x250um.png}
\end{subfigure}
\begin{subfigure}[b]{.49\textwidth}
\includegraphics[width=\textwidth]{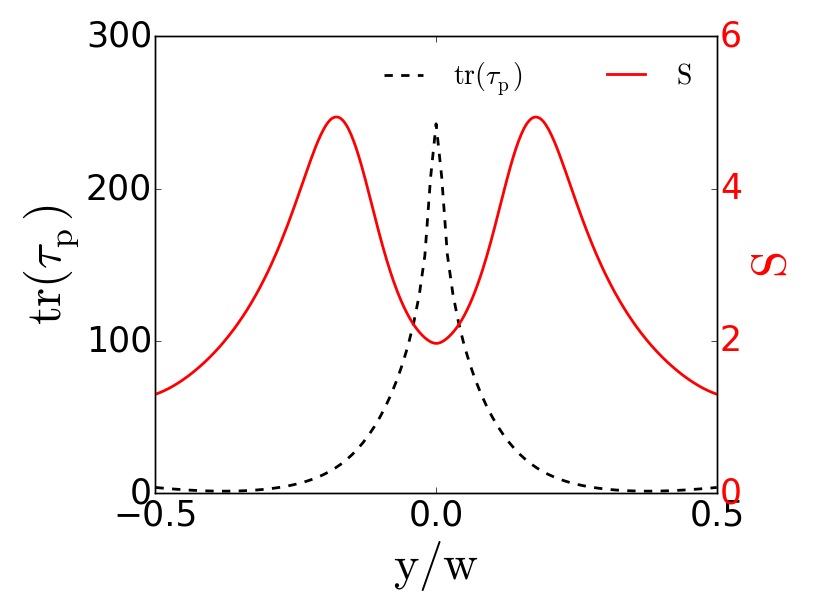}
\caption{}
\label{stress_stretching_u600ums_x250um.png}
\end{subfigure}
\caption{(a) Flow-type parameter ($\rm{\Lambda}$) and local Wi ($\rm{Wi_{local}}$), and (b) the trace of polymeric stress ($\rm{tr(\tau_p)}$) and stretching ($S$) along the width of cross-slot geometry close to stagnation point (along the white dashed line shown in Fig. \ref{cross_slot_strong_strain_rate.png}a(ii)) at $\rm{Wi=0.6}$.}
\end{figure}

Our analysis of ordered viscoelastic flows establishes a general relationship between ${\rm tr}(\boldsymbol{\tau_p})$ and $S$ for different steady and weakly unsteady flows at $\rm{Wi \ll 1}$ (Eq. \ref{stretching_stress_relation_unsteady_si}). Further, the analysis of flows undergoing \textit{motions with constant stretch history} (MWCSH) uncovers the relationships between ${\rm tr}(\boldsymbol{\tau_p})$ and $S$ at $\rm{Wi \gg 1}$ for purely shear (Eq. \ref{stretching_stress_relation_strong_flow_shear_si}) and purely extensional (Eq. \ref{stretching_stress_relation_strong_flow_lin_ext_si}) flows. The stress-stretching relationship for pure extensional flows (Eq. \ref{stretching_stress_relation_strong_flow_lin_ext_si}) is different than pure shear flows (Eq. \ref{stretching_stress_relation_strong_flow_shear_si}) at $\rm{Wi \gg 1}$, and it is not possible to analytically derive either the stress field or the stretching field at $\rm{Wi>1}$ for flows having mixed-kinematics (i.e., non-linear extensional flows). Thus, we used numerical simulations to examine the relationship between stress and stretching fields for mixed-kinematics at ${\rm Wi} \gtrsim O(1)$. We numerically calculated the stress and stretching fields for viscoelastic flow in a cross-slot geometry for small ($\rm{Wi=0.01}$), moderate ($\rm{Wi=0.6}$), and large ($\rm{Wi=4}$) Weissenberg numbers (Fig. \ref{cross_slot_strong_strain_rate.png}). 
 We have also calculated the flow-type parameter, $\Lambda$, and the local Wi ($\rm{Wi_{local}= \lambda \lvert\dot{\gamma}\rvert}$) to determine the local flow kinematics and strength, respectively. 
 At $\rm{Wi \ll 1}$, we demonstrated in the main text that there exists a direct relationship between the stress field and stretching field (Eq. \ref{stretching_stress_relation_unsteady_si}). 
 This result is supported by simulations (Fig. \ref{cross_slot_strong_strain_rate.png}(i)) and due to weak viscoelasticity, defined here as $\Lambda {\rm Wi_{local}}$ (Fig. \ref{cross_slot_strong_strain_rate.png}b(i)). Therefore, at $\rm{Wi=0.01}$, the stress field and stretching field are highly correlated (Fig. \ref{cross_slot_strong_strain_rate.png}(i)). At moderate Wi ($\rm{Wi=0.6}$), prior to viscoelastic instability, the flow is still symmetric in the cross-slot geometry. However, in the extension-dominated region, a streak of high polymeric stress forms (Fig. \ref{cross_slot_strong_strain_rate.png}c(ii)), which does not coincide with the regions of high stretching (Fig. \ref{cross_slot_strong_strain_rate.png}d(ii)). At $\rm{Wi=0.6}$, the extension is non-uniform (i.e, mixed-kinematics) (Fig. \ref{cross_slot_strong_strain_rate.png}a(ii)), and the flow exhibits strong deformation (${\rm Wi_{local}} \sim O(1)$; Fig. \ref{cross_slot_strong_strain_rate.png}b(ii)). The flow-type parameter ($\Lambda$) and the local Wi ($\rm{Wi_{local}}$) along the width of the cross-slot geometry close to the stagnation point at $\rm{Wi=0.6}$ are shown in Fig. \ref{flow_type_Wi_local_u600ums_x250um.png}. The value of $\Lambda$ varies in the range of $0<\Lambda \leq 1$ indicating mixed-kinematics and $\rm{Wi_{local}>1}$ represents strong deformation (Fig. \ref{flow_type_Wi_local_u600ums_x250um.png}). 
 Examining the stress and stretching along the width of the geometry at $\rm{Wi=0.6}$ (Fig. \ref{stress_stretching_u600ums_x250um.png}) reveals that the relationship between the stress field and stretching field breaks down for strong flows ($\rm{Wi_{local} \gtrsim O(1)}$) having mixed-kinematics. However, at $\rm{Wi=4}$, the strong correlation between the stretching and stress fields is recovered, whereby the streaks of the polymeric stress field once again coincide with the ridge of the stretching field (Fig. \ref{cross_slot_strong_strain_rate.png}(iii)). At large Wi, the flow inside the cross-slot geometry becomes asymmetric due to elastic instability \cite{Arratia2006}, and the region where the streaks of polymeric stress form becomes shear-dominated (Fig. \ref{cross_slot_strong_strain_rate.png}a(iii)). Our theoretic analysis of MWCSH flows has shown that there exists a direct relationship between the stress and stretching for both homogeneous and non-homogeneous shear flows at $\rm{Wi \gg 1}$ (Eq. \ref{stretching_stress_relation_strong_flow_shear_si}). Therefore, again we see a strong correlation between the stress and stretching fields at large Wi. 
 For all the geometries explored in the present study, the regions where  streaks of polymeric stress form become shear-dominated because these streaks act like barriers that resist the flow crossing the region of high stress (Figs. \ref{flow_type_simple_geometry.png} and \ref{flow_type_complex_geometry.png}). Therefore, there is very strong correlation between the stress and stretching fields specifically in the region of high stress. 
 
\subsection*{Effect of integration time on stretching field}
The main text primarily focuses on linking the polymeric stress, ${\rm tr}(\boldsymbol{\tau_p})$, to the Lagrangian stretching, $S$, accrued over the relaxation time of the polymer ($\Delta t = \lambda$). 
For small Wi, the topology of $S$ is expected to remain unchanged with varying integration time ($\Delta t$). 
However, the magnitude of the stretching increases with $\Delta t$.  
For $\rm{Wi \ll 1}$, the relationship between ${\rm tr}(\boldsymbol{\tau_p})$ and $S$ for arbitrary integration time is:
\begin{equation}\label{stretching_stress_relation_dt}
{\rm tr}(\boldsymbol{\tau_p})= \frac{8}{n^2m^2}\frac{(b_{11}-b_2)}{\lambda^2}g(t_0)(S^n-1)^2,
\end{equation}
where $\Delta t= m\lambda, m>0$. Similarly, the corresponding relationships at $\rm{Wi \gg 1}$ for purely shear and purely extensional flows are: 
\begin{equation}\label{stretching_stress_relation_strong_flow_shear _dt}
{\rm tr}(\boldsymbol{\tau_p})=2\frac{(b_{11}-b_2)}{\lambda^2 m^2}g(t_0)S^2,
\end{equation}
 and
\begin{equation}\label{stretching_stress_relation_strong_flow_lin_ext_dt}
{\rm tr}(\boldsymbol{\tau_p})=2\frac{(b_{11}-b_2)}{\lambda^2m^2}g(t_0) (ln(S))^2,
\end{equation}
respectively.

\begin{figure}[!ht]
 \centering
 \includegraphics[width=\textwidth]{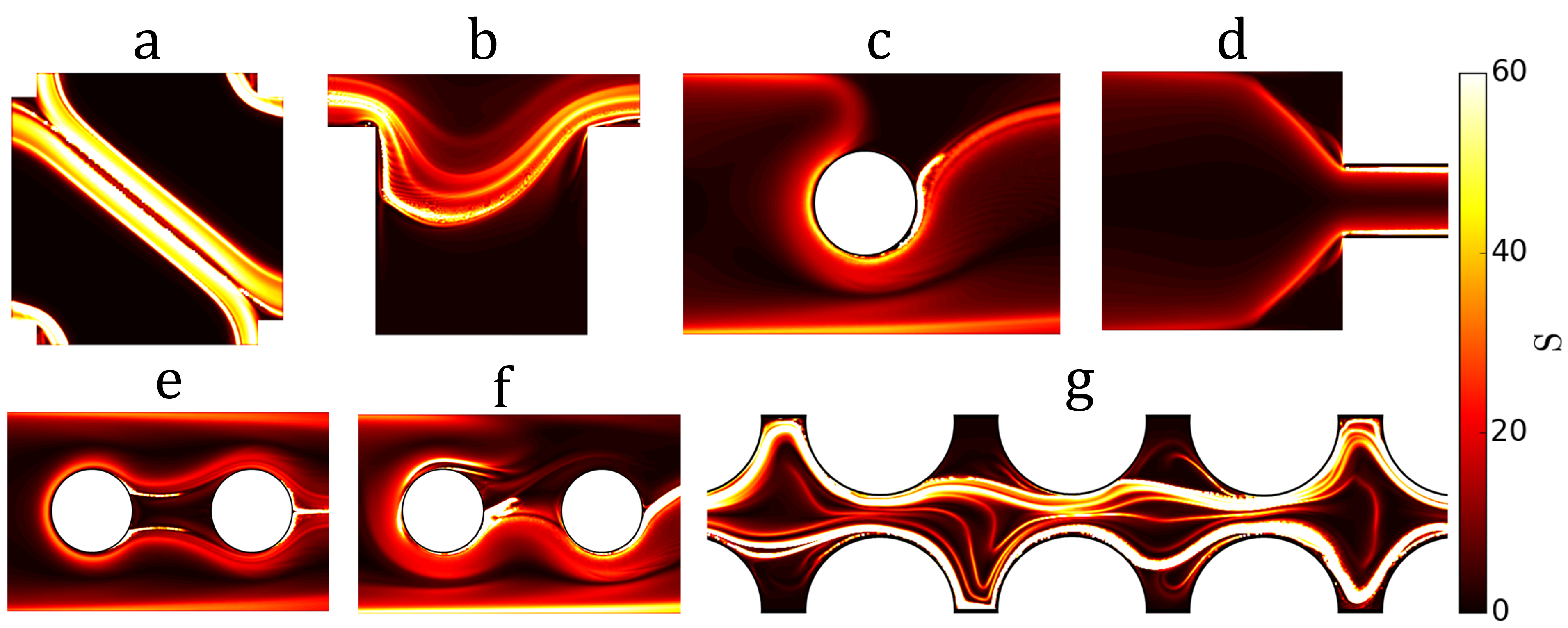}
 \caption{\label{stretching_dt_2ld_all_geometries.png} Stretching fields obtained using integration time $\Delta t=2\lambda$ for viscoelastic flows in various geometries (see also main text Figs.~1~and~3) at large Weissenberg number (${\rm Wi} \gtrsim 1$): (a) cross-slot geometry at ${\rm Wi}=4$, (b) flow over a cavity at ${\rm Wi}=1.25$, (c) cylinder confined in a channel at ${\rm Wi}=2.5$, (d) flow through an isolated constriction at ${\rm Wi}=0.75$, (e) two streamwise cylinders in a channel at moderate Wi (${\rm Wi}=1.88$),  (f) two streamwise cylinders in a channel at large Wi (${\rm Wi}=3.12$), and
 (g) corrugated channel at ${\rm Wi}=1.68$. }
 \end{figure}

For the moderate to large Wi (${\rm Wi} \gtrsim 1$) investigated through numerical simulation in the present work, we examine the sensitivity of the stretching field to $\Delta t$.
Stretching fields for all geometries presented in the main text (Figs.~1~and~3) were additionally computed for an integration time of twice the polymer relaxation time ($\Delta t=2\lambda$) for comparison (Fig.~\ref{stretching_dt_2ld_all_geometries.png}).
In the stretching fields of chaotic flows, secondary ridges and finer structures are further refined as the value of integration time increases (Figs. \ref{stretching_dt_2ld_all_geometries.png}b and \ref{stretching_dt_2ld_all_geometries.png}g). 
However, the stretching manifolds (i.e. ridges of maximal stretching), which are primarily responsible for controlling the flow dynamics, remain nearly unchanged in all geometries (see also Figs.~1~and~3 in the main text).

\subsection*{Cross-correlation of stretching and stress fields}
\begin{figure}[!ht]
 \centering
 \includegraphics[width=\textwidth]{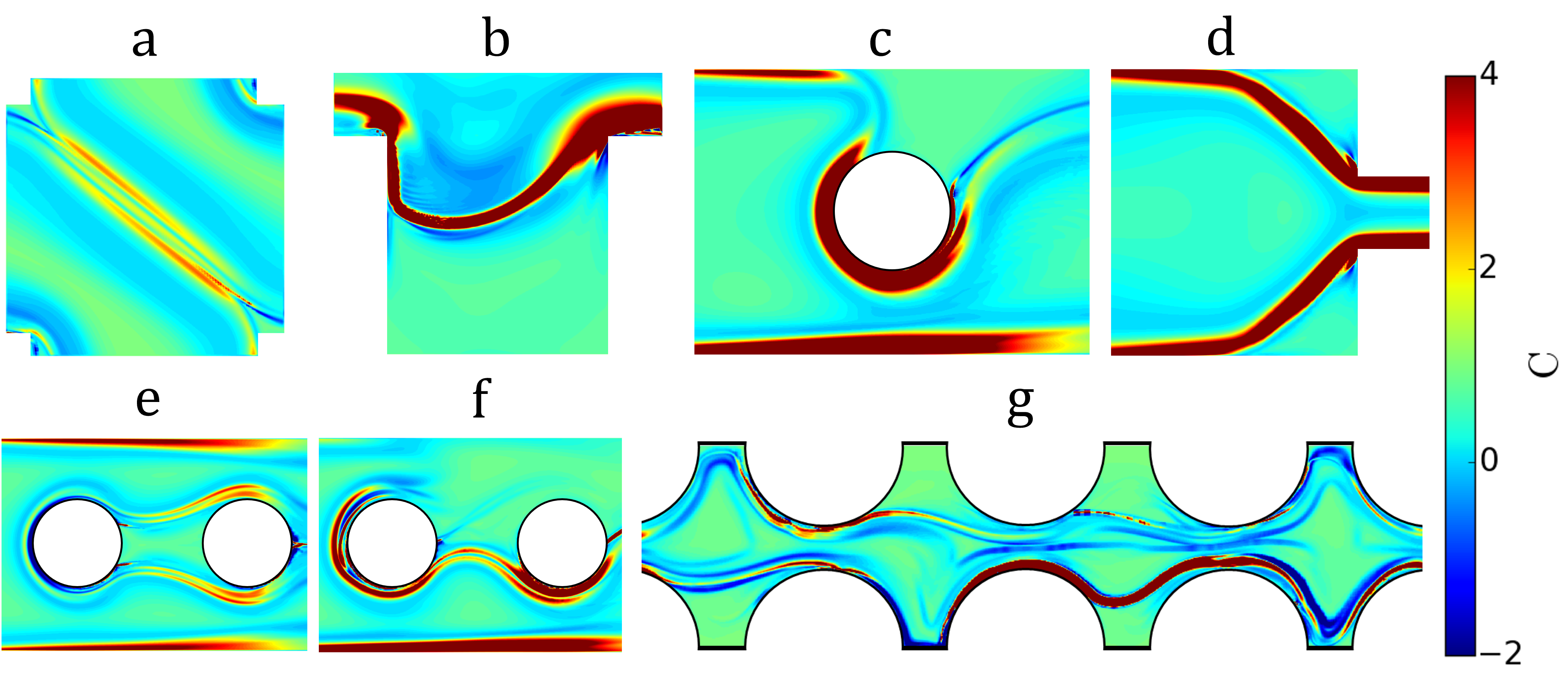}
 \caption{\label{cross_correlation_all_geometries.png} Cross-correlation between the stress field and the stretching field for viscoelastic flows in different geometries (see main text Figs.~1~and~3) at large Weissenberg number (${\rm Wi} \gtrsim 1$): (a) cross-slot geometry at ${\rm Wi}=4$, (b) flow over a cavity at ${\rm Wi}=1.25$, (c) cylinder confined in a channel at ${\rm Wi}=2.5$, (d) flow through an isolated constriction at ${\rm Wi}=0.75$, (e) two streamwise cylinders in a channel at moderate Wi (${\rm Wi}=1.88$),  (f) two streamwise cylinders in a channel at large Wi (${\rm Wi}=3.12$), and
 (g) corrugated channel at ${\rm Wi}=1.68$. }
 \end{figure}

As an additional comparison between the topologies of the stress and stretching field, we compute the local cross-correlation, $\rm{C}$, between these fields for the flows presented in the main text (Figs.~1~and~3).
The cross-correlation function is defined as:
\begin{equation}\label{cross_correlation}
    \rm{C}= \frac{[ {\rm tr}(\boldsymbol{\tau_p}) - \langle {\rm tr}(\boldsymbol{\tau_p}) \rangle ] \cdot [S- \langle S \rangle] }
    {\langle {\rm tr}(\boldsymbol{\tau_p}) \rangle \langle S \rangle},
\end{equation}
where $\langle \cdot \rangle$ represents the mean value. The strength of the cross-correlation for different geometries are shown in Fig. \ref{cross_correlation_all_geometries.png}. A large cross-correlation value indicates similarly high local values of both fields (i.e. positive correlation). 
For all the geometries examined, there exists a strong correlation between the polymeric stress field and stretching field, especially in the regions of observed high stress and stretching (Fig. \ref{cross_correlation_all_geometries.png}). The ridges characterized by high stress (or stretching) ultimately control the flow states and dynamics in these viscoelastic flows.  

\subsection*{Stretching and stress in 3D geometry}
\begin{figure}[!ht]
 \centering
 \includegraphics[width=\textwidth]{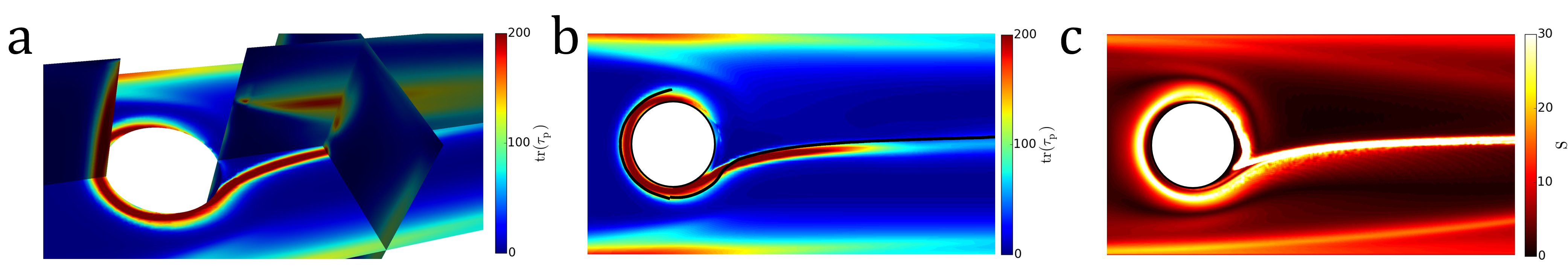}
 \caption{\label{single_cylinder_u400ums_3D_tau_s.png} 
  Viscoelastic flow around a cylinder confined in a 3D channel at ${\rm Wi}=2.5$: (a) isometric view representing the trace of the polymeric stress tensor (${\rm tr}(\boldsymbol{\tau_p})$) on different planes of the 3D geometry, (b) ${\rm tr}(\boldsymbol{\tau_p})$ from the mid-plane of the channel , and (c) stretching field for the same plane shown in (b). The black line (b) represents the stretching manifold corresponding to the material line of maximum value of the stretching field in (c).}
 \end{figure}
 
All comparisons between polymeric stress and Lagrangian stretching in the main text focused on two-dimensional (2D) geometries. 
To illustrate that the correlation between stress and and stretching persists in three-dimensional (3D) viscoelastic flows, we also perform a 3D numerical simulation for a channel having a cylindrical obstruction (Fig. \ref{single_cylinder_u400ums_3D_tau_s.png}; comparable to the 2D geometry in Fig.~1c in the main text). 
The polymeric stress field and stretching field on the mid-plane normal to the axis of cylinder are shown in Fig. \ref{single_cylinder_u400ums_3D_tau_s.png}b and Fig. \ref{single_cylinder_u400ums_3D_tau_s.png}c, respectively. Further, we extract the line corresponding to the maximum stretching within that plane, which is superimposed on the stress field (Fig. \ref{single_cylinder_u400ums_3D_tau_s.png}b). 
The fields are highly correlated and the line of maximum stretching coincides with the streak of stress field, as observed in 2D (see main text Figs.~1~and~3).

\subsection*{sPTT and Giesekus models of viscoelastic flow}

 \begin{figure}[!ht]
 \centering
 \includegraphics[width=0.9\textwidth]{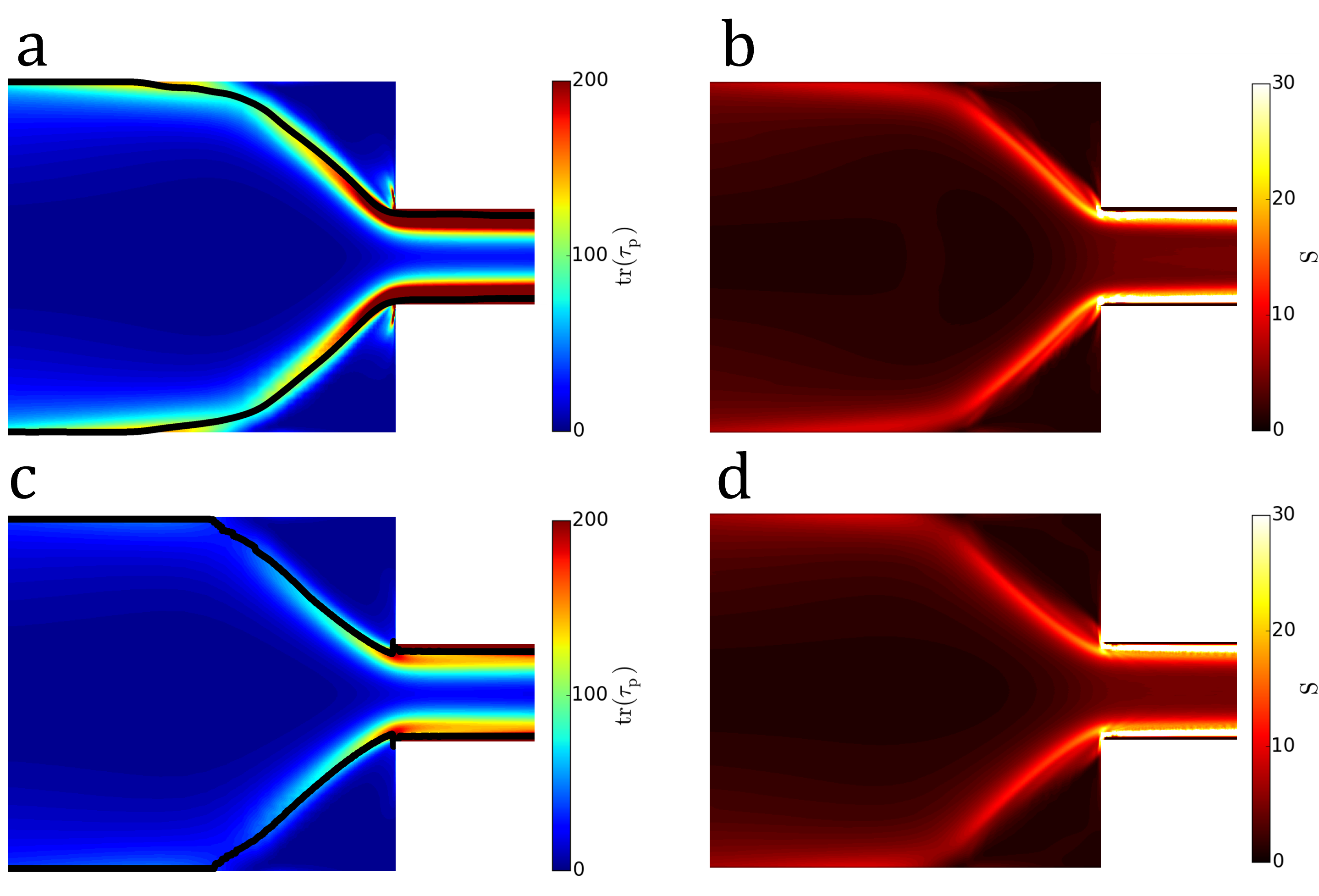}
 \caption{\label{constriction_flow_ptt.png}  (a,c) Polymeric stress field and (b,d) stretching field for viscoelastic flow through a sudden constriction (see also main text Fig.~1d) at $\rm{Wi=0.75}$ using (a,b) linear sPTT model and (c,d) Giesekus model. 
 The values of model parameters used for sPTT model are $\beta=0.05$, $\lambda =1$ s, and $\epsilon=0.001$. 
 The parameters used for Giesekus model are $\beta=0.05$, $\lambda =1$ s, and $\alpha=0.02$. The black lines (a,c) represent the lines of maximum stretching.}
 \end{figure}

The primary results of our numerical simulations (presented in the main text), rely on the widely used FENE-P constitutive model. 
To determine the sensitivity of our results to the rheological fluid model, we performed additional simulations for the flow-through constriction geometry (see main text Fig.~1d) using the linear sPTT and Giesekus models of viscoelastic fluid. 
The results for the sPTT and Giesekus models confirm our FENE-P results and show a very high correlation between the trace of the polymeric stress field (Figs. \ref{constriction_flow_ptt.png}a,c) and the stretching field (Figs. \ref{constriction_flow_ptt.png}b,d). 
Likewise, the stretching manifolds (ridge of maximum stretching) extracted from the stretching field coincide with the streaks of the polymeric stress field to a similar degree across all constitutive models (Figs. \ref{constriction_flow_ptt.png}a,c).

\subsection*{Time-dependent chaotic flow through corrugated channel}
\begin{figure}[!ht]
 \centering
 \includegraphics[width=\textwidth]{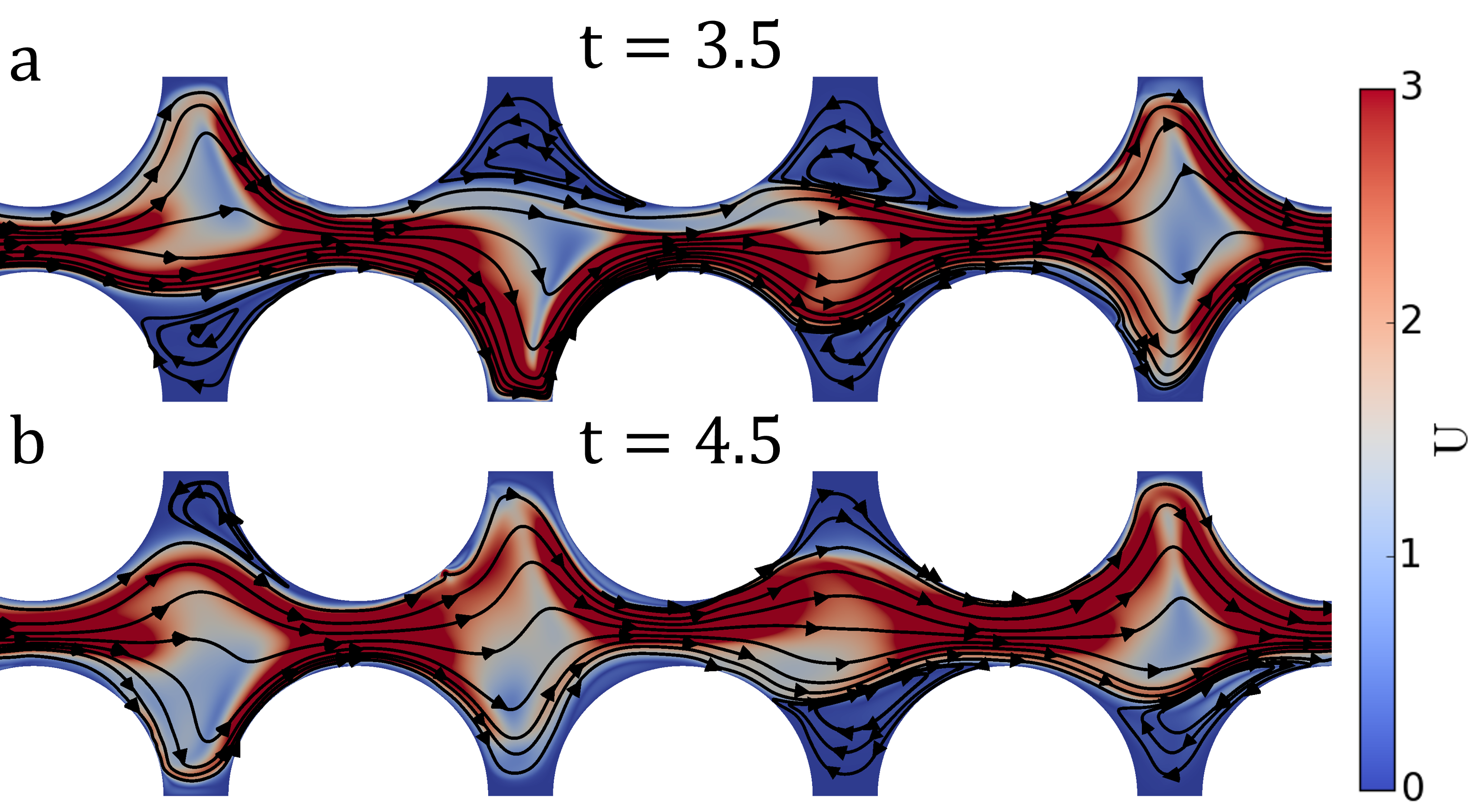}
 \caption{\label{multistability_velocity_fm179_225.png} Instantaneous velocity fields for viscoelastic flow through corrugated channel (corresponding to Fig.~3c) at (a) $\rm t \approx 3.5$ and (b) $\rm t \approx 4.5$. Time, $\rm  t$, has been normalized with polymeric relaxation time and $\rm Wi=1.68$.}
 \end{figure}
To emphasize the time-dependent nature of the viscoelastic flows examined in the main text, we provide an example for the chaotic flow through the corrugated channel (main text Fig.~3) \cite{Kumar2021multistability}. 
The velocity fields for this flow at different instances of time are shown in Fig.~\ref{multistability_velocity_fm179_225.png}, which illustrates that the flow states across the various pores readily evolve in time (see also supplementary movie-4 of Ref. \cite{Kumar2021multistability} for this same data set). 

\section*{Quantitative relation between stress and stretching fields for a simulated geometry}
\begin{figure}[!ht]
\centering
\begin{subfigure}[b]{0.32\textwidth} 
\includegraphics[width=\textwidth]{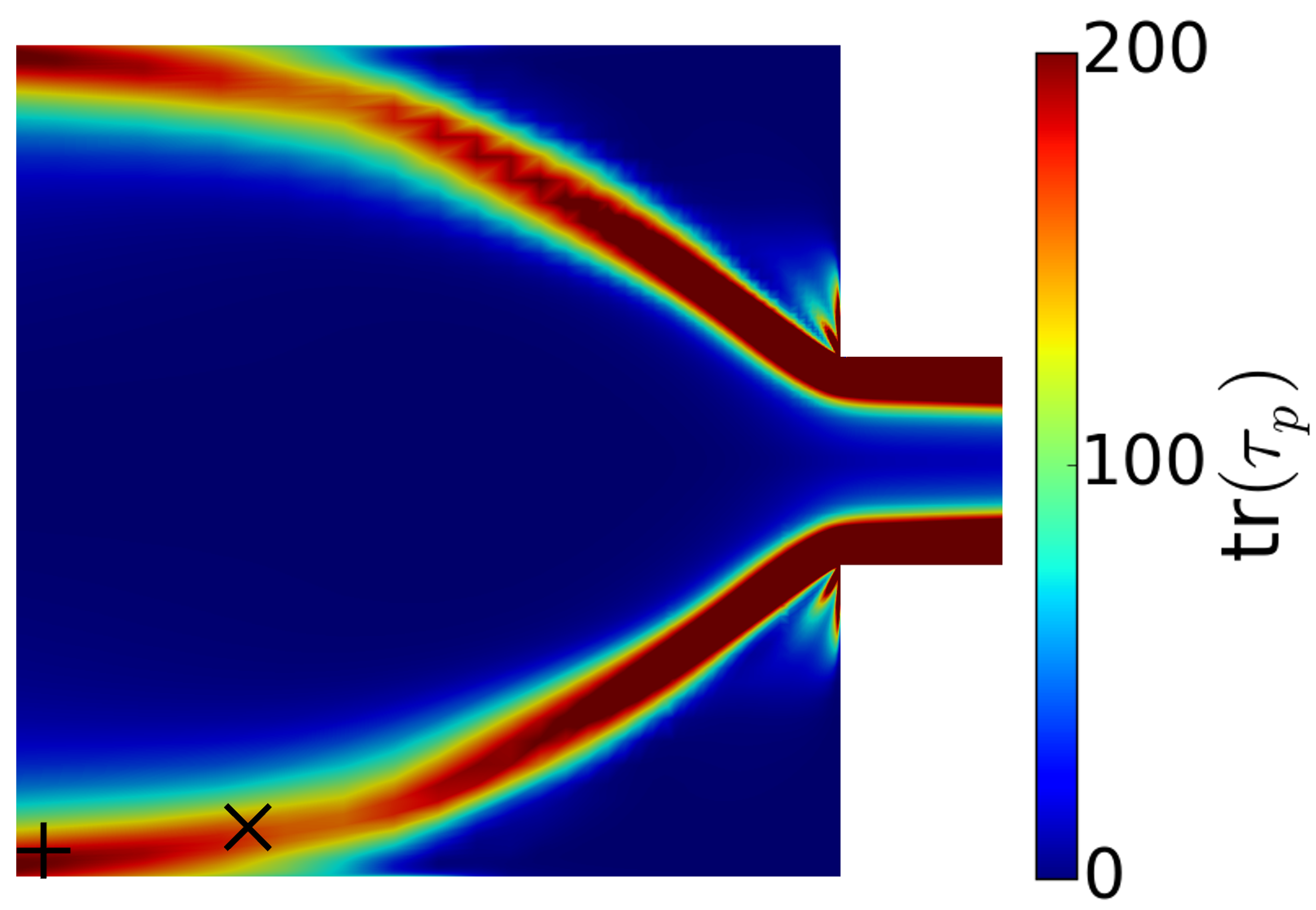}
\caption{}
\label{stress_constriction_u1mms.png}
\end{subfigure}
\begin{subfigure}[b]{.32\textwidth}
\includegraphics[width=\textwidth]{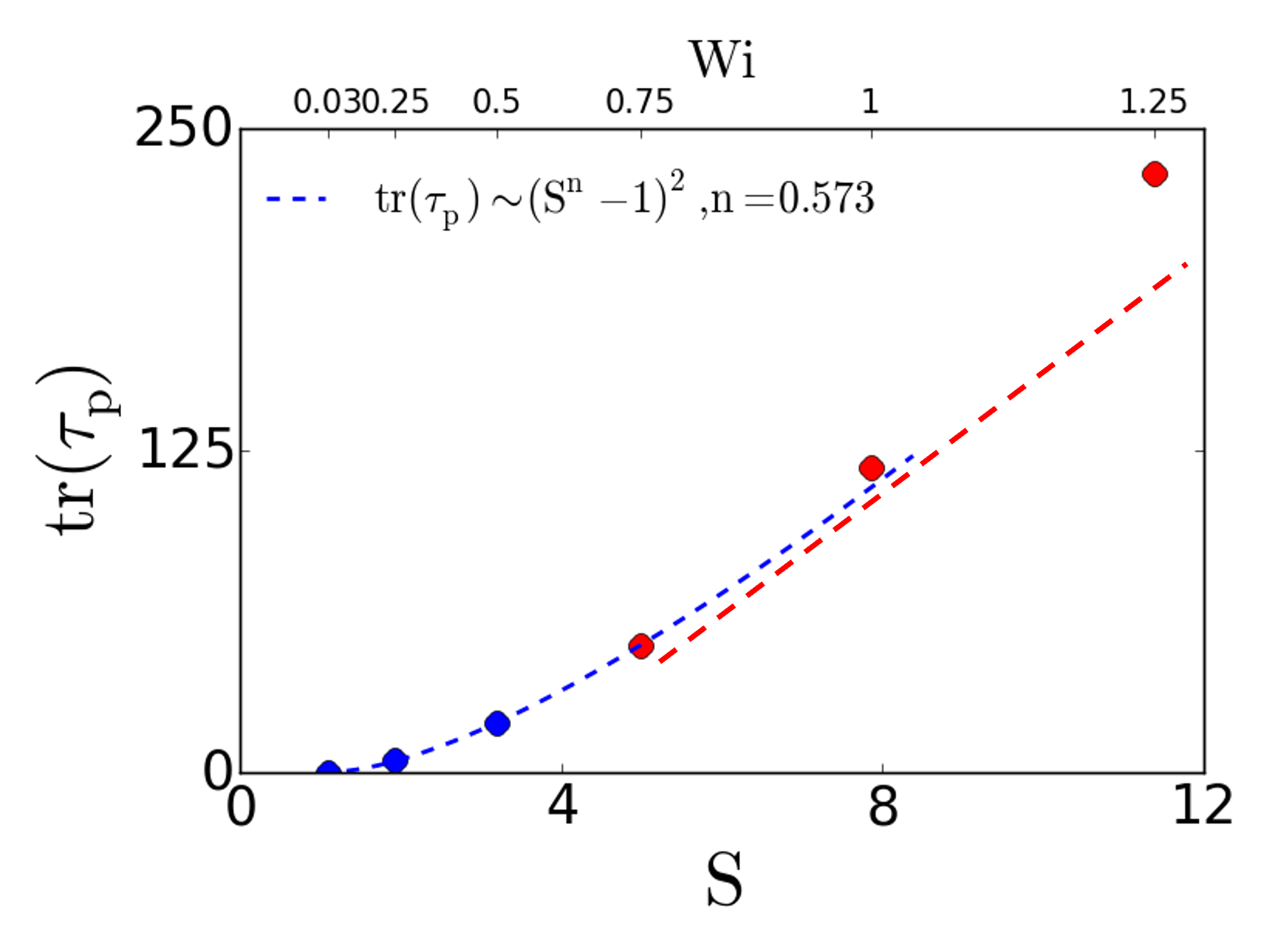}
\caption{}
\label{trtau_s_local_x1mm_370um.png}
\end{subfigure}
\begin{subfigure}[b]{.32\textwidth}
\includegraphics[width=\textwidth]{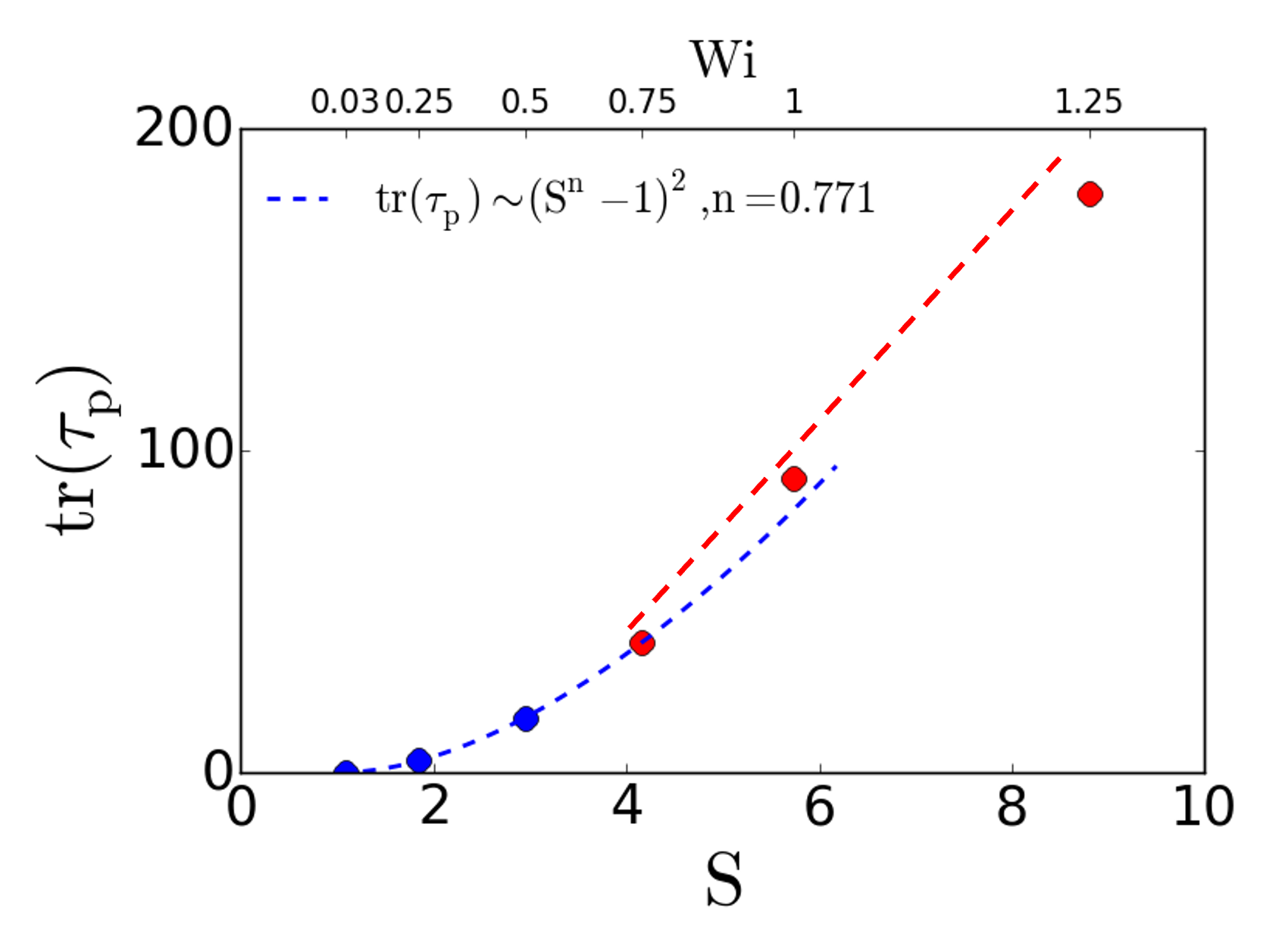}
\caption{}
\label{trtau_s_local_x700um_340um.png}
\end{subfigure}
\caption{\textcolor{black}{Polymeric stress field for a viscoelastic flow through a constriction at $\rm{Wi=1.25}$. The local values of the polymeric stress and stretching fields for different Wi at the locations indicated by (b) $+$ and (c) $\times$ in Fig. \ref{stress_constriction_u1mms.png}.}}
\label{local_tau_s}
\end{figure}

\textcolor{black}{While the main text primarily focuses on the topological similarity between the polymeric stress field and the stretching field, the analytical results suggest the possibility of establishing a quantitative relationship between the local values of the stress and stretching fields. The evolution of the local values of the polymeric stress and stretching fields with Wi at different locations for a viscoelastic flow through the constriction are shown in Fig. \ref{local_tau_s}. The local polymeric stress follows the analytically predicted scaling ${\rm tr(\tau_p)} \sim (S^n-1)^2 $ as $\rm Wi \to 0$ (Figs. \ref{trtau_s_local_x1mm_370um.png} and \ref{trtau_s_local_x700um_340um.png}). These results provide a significant foundation to further investigate the dependency of the scaling exponent $n$ and the relation between the polymeric stress and stretching at large Wi in mixed flows.}

\bibliography{apssamp}

\end{document}